\documentclass[aps,prapplied,superscriptaddress,twocolumn,floatfix,10pt]{revtex4-2}

\usepackage{amsmath}
\usepackage{iftex}
\ifPDFTeX
  \usepackage[utf8]{inputenc}
  \usepackage[T1]{fontenc}
  \usepackage{newtxtext}
\else
  \usepackage{fontspec}
\fi
\usepackage{newtxmath}

\usepackage{graphicx}
\usepackage{booktabs}
\usepackage[dvipsnames]{xcolor}
\usepackage[normalem]{ulem}
\usepackage{chemformula}
\usepackage{placeins,etoolbox}
\preto\subsection{\FloatBarrier}

\usepackage[colorlinks=true,citecolor=blue,linkcolor=blue,urlcolor=blue]{hyperref}
\setcitestyle{super}

\graphicspath{{figures/}}

\begin{document}

\title{Mn--Cr dopant association in dilute co-doped \ch{ZnAl2O4} revealed by combined EPR and optical spectroscopy}

\author{Apoko~S.~Omondi}
\affiliation{HUN-REN Wigner Research Centre for Physics, Institute for Solid State Physics and Optics, P.O.\ Box 49, H-1525 Budapest, Hungary}
\affiliation{Department of Physical Chemistry and Materials Science, Faculty of Chemical Technology and Biotechnology, Budapest University of Technology and Economics, M\H{u}egyetem rakpart 3, H-1111 Budapest, Hungary}

\author{M\'aty\'as~M.~Rudolf}
\affiliation{HUN-REN Wigner Research Centre for Physics, Institute for Solid State Physics and Optics, P.O.\ Box 49, H-1525 Budapest, Hungary}
\affiliation{Department of Physical Chemistry and Materials Science, Faculty of Chemical Technology and Biotechnology, Budapest University of Technology and Economics, M\H{u}egyetem rakpart 3, H-1111 Budapest, Hungary}

\author{Nikoletta~Jegeny\'es}
\affiliation{HUN-REN Wigner Research Centre for Physics, Institute for Solid State Physics and Optics, P.O.\ Box 49, H-1525 Budapest, Hungary}

\author{\'Arp\'ad~Jakab}
\affiliation{HUN-REN Wigner Research Centre for Physics, Institute for Solid State Physics and Optics, P.O.\ Box 49, H-1525 Budapest, Hungary}

\author{G\'abor~Bortel}
\affiliation{HUN-REN Wigner Research Centre for Physics, Institute for Solid State Physics and Optics, P.O.\ Box 49, H-1525 Budapest, Hungary}

\author{Bence~G.~M\'arkus}
\affiliation{Stavropoulos Center for Complex Quantum Matter, Department of Physics and Astronomy, University of Notre Dame, Notre Dame, Indiana 46556, USA}

\author{Adam~Gali}
\affiliation{HUN-REN Wigner Research Centre for Physics, Institute for Solid State Physics and Optics, P.O.\ Box 49, H-1525 Budapest, Hungary}
\affiliation{Department of Atomic Physics, Institute of Physics, Budapest University of Technology and Economics, M\H{u}egyetem rakpart 3, H-1111 Budapest, Hungary}
\affiliation{MTA-WFK Lend\"ulet ``Momentum'' Semiconductor Nanostructures Research Group, P.O.\ Box 49, H-1525 Budapest, Hungary}

\author{D\'avid~Beke}
\email{beke.david@wigner.hun-ren.hu}
\affiliation{HUN-REN Wigner Research Centre for Physics, Institute for Solid State Physics and Optics, P.O.\ Box 49, H-1525 Budapest, Hungary}
\affiliation{Kand\'o K\'alm\'an Faculty of Electrical Engineering, \'Obuda University, Tavaszmez\H{o} 17, H-1084 Budapest, Hungary}

\begin{abstract}
Efficient and durable phosphors made from abundant elements underpin energy-saving solid-state lighting, near-infrared light sources for sensing, and X-ray and biomedical imaging. Co-doping is a standard route to brighter phosphors, yet the fate of the co-dopants is rarely established. We combine quantitative electron paramagnetic resonance (EPR) spin counting with photoluminescence, photoluminescence excitation, and X-ray-excited spectroscopy across two dilute Mn/Cr co-doping series of hydrothermally grown \ch{ZnAl2O4} nanoparticles and show that the two ions can form an exchange-coupled, edge-sharing Cr--Mn defect complex, with \ch{Mn^{2+/4+}} entering an octahedral site next to \ch{Cr^3+}. The octahedral Mn lifts the inversion symmetry of the \ch{Cr^3+} site and strengthens its parity-forbidden absorption, enhancing the \ch{Cr^3+} emission without energy transfer. The population of the defect complex follows the law of mass action, and above a threshold Mn concentration, Mn re-partitions into EPR-silent octahedral configurations that quench both emissions. Concentration optima in co-doped phosphors can thus reflect defect speciation, not only energy-transfer efficiency, which opens a structural design handle for brighter emitters.
\end{abstract}

\maketitle

\noindent\textbf{Keywords:} Zinc aluminate, EPR, dopant association, exchange coupling, sensitization, XEOL, photoluminescence

\section{Introduction}

Inorganic phosphors, which convert ultraviolet or blue light and ionizing radiation into visible and near-infrared (NIR) emission, are embedded in technologies that shape everyday life. Phosphor-converted white light-emitting diodes (LEDs) have transformed general lighting, a sector that accounts for a major share of electricity consumption,\cite{Pimputkar2009} and the efficacy and color quality of these devices depend critically on efficient red-emitting phosphors.\cite{Pust2014,Zhu2014} Deep-red and NIR-emitting phosphors in turn enable compact broadband light sources for portable spectroscopy, food and health monitoring, and night vision,\cite{Rajendran2018,Jia2020} as well as persistent luminescent probes for in vivo imaging\cite{Maldiney2014} and nanoscintillators for low-dose, high-resolution X-ray imaging.\cite{ChenQ2018,Ou2021} Across these applications, the same requirements recur: high emission efficiency, chemical and thermal stability, and, for deployment at scale, hosts composed of abundant, inexpensive elements. Meeting them requires control not only over which activator is introduced, but also over where it sits in the lattice and with which other defects it interacts.

Zinc aluminate (\ch{ZnAl2O4}, ZAO), a wide-band-gap oxide of the spinel family \ch{AB2O4}, is a chemically and thermally robust phosphor host with the potential to meet these criteria. It emits both intrinsically and through a variety of dopant-related luminescent centers.\cite{Cornu2013,Verger2016,Pathak2018,Yang2023} When activated with transition-metal or rare-earth ions, its emission can be tuned across a broad spectral range through the choice of dopant, its concentration, and the lattice site it occupies. Governed by charge compensation,\cite{Khaidukov2020} ionic radius, and synthesis conditions, dopants can enter either the tetrahedral or the octahedral sublattice, giving rise to a rich, site-dependent optical response. ZAO is moreover composed of abundant, inexpensive elements, unlike its gallate analog \ch{ZnGa2O4}, which makes it an attractive host for large-scale applications. However, the luminescence of ZAO often falls short of that of other spinels and of competing phosphor and scintillator materials, and considerable effort has therefore been devoted to understanding and improving its optical performance. Here, the interaction of luminescent centers with the host, with other dopants, and with unintentionally created point defects is central. It is especially important in spinels, in which the cations can occupy different sites and the site distribution shapes the overall material properties, including the optical response.

One of the most commonly used strategies to increase quantum yield is co-doping. When two different impurities coexist in the same host, interionic interactions can modify the magneto-optical properties of one or both centers. In ZAO, such interactions have mainly been exploited for sensitization, whereby a donor ion whose emission band overlaps with the excitation band of the acceptor is introduced to enhance acceptor luminescence.\cite{Lakshminarayana2011} This strategy can compensate for the weak absorption of activators whose excitation relies on spin-forbidden transitions\cite{Xiao2025} and can steer the relaxation dynamics toward the desired emission. Sensitization is particularly attractive for ZAO, where \ch{Cr^3+} is a notably inefficient emitter, in contrast to the gallate analog \ch{ZnGa2O4}.\cite{Basavaraju2015}

\ch{Cr^3+} and \ch{Mn^2+} constitute such a donor--acceptor pair in this host. Manganese stabilizes in the spinel lattice in several oxidation states, most commonly \ch{Mn^2+} and \ch{Mn^4+}, both of which are magnetically active, and can occupy either the A or the B site, with optical properties dictated by the local crystal field.\cite{Song2014} In the tetrahedral site, \ch{Mn^2+} emits green light usually centered at around 510~nm through the $^{4}T_{1}(^{4}G) \rightarrow {}^{6}A_{1}$ transition, whereas in the strong-field octahedral site its emission shifts to the red, near 650~nm. \ch{Mn^4+}, in contrast, occupies the octahedral site almost exclusively, substituting for \ch{Al^3+}, and emits deep-red light around 678~nm via the $^{2}E_{g} \rightarrow {}^{4}A_{2g}$ transition.\cite{Trung2020} \ch{Cr^3+} substitutes exclusively for \ch{Al^3+} in the octahedral site, where its $^{2}E_{g} \rightarrow {}^{4}A_{2g}$ emission comprises the sharp zero-phonon R and N lines together with their phonon sidebands. The \ch{Mn^2+} $^{4}T_{1}(^{4}G) \rightarrow {}^{6}A_{1}$ emission band overlaps the \ch{Cr^3+} $^{4}A_{2g} \rightarrow {}^{4}T_{2g}$ absorption band at 520--530~nm. Excited \ch{Mn^2+} can therefore, in principle, relax non-radiatively by populating \ch{Cr^3+} excited states, whose radiative decay then enhances the red emission.

Indeed, Mn--Cr energy transfer has repeatedly been used to explain \ch{Cr^3+} luminescence enhancement in co-doped systems.\cite{Zhu2021,Kim2004,Hou2026,Hou2025} The rate of such resonant transfer, however, falls off with the sixth power of the donor--acceptor distance, making it effective only for closely spaced Mn--Cr pairs. Zhu et al., for example, attributed the \ch{Cr^3+} photoluminescence enhancement in Mn/Cr co-doped spinels to multipolar Mn$\rightarrow$Cr energy transfer, yet they also observed the average \ch{Cr^3+} lifetime to decrease with increasing Mn concentration --- contrary to what is expected for the acceptor in an energy-transfer process.\cite{Zhu2021,Chen2023,Su2018} Such inconsistencies suggest that energy transfer alone does not fully account for the Mn--Cr interaction and that additional processes, invisible to optical spectroscopy, may be at work.

In this work, we combine optical spectroscopy with X-band electron paramagnetic resonance (EPR) to resolve the dilute Mn--Cr interaction in the ZAO host. EPR is sensitive to the local symmetry and environment of paramagnetic centers and thus offers a luminescence-independent view of the origin of the optical response.\cite{Zhang2016} We synthesized two series of ZAO nanoparticles with fixed Cr and varied Mn concentrations and characterized their structure and morphology. The photoluminescence (PL), photoluminescence excitation (PLE), X-ray-excited optical luminescence (XEOL), and EPR data were then combined and correlated using statistical methods to characterize the interaction between the two impurity centers.

We found that Mn co-doping enhances the \ch{Cr^3+} emission, with an optimal Mn concentration that scales with the Cr content, even though the system operates outside the resonant energy-transfer regime. The enhancement instead points to the association of the two dopants into exchange-coupled Mn--Cr defect complexes within the lattice. These results suggest that, in dilute co-doped phosphors, dopant association rather than energy transfer alone can set the attainable brightness.

Because this conclusion rests on correlating several techniques, we keep the evidence from each technique separate from its joint interpretation. The \hyperref[sec:results]{Results and discussion} section presents, technique by technique, the measured data, their evaluation, and the conclusions that each technique supports on its own. The \hyperref[sec:discussion]{Discussion} section then combines these observations, together with their statistical correlation, into a single model of the Mn--Cr association. Synthesis, measurement details, and the statistical procedures are described in the \hyperref[sec:experimental]{Experimental section}.

\section{Results and discussion}
\label{sec:results}

\subsection{Material synthesis}

Two series of ZAO containing 0.04 and 0.32~mol\% Cr (hereafter ZAO4 and ZAO32) were synthesized via a hydrothermal route and subsequently annealed at 1100~$^{\circ}$C. The procedure is detailed in the \hyperref[sec:methods]{Materials and methods} section. In each series, the Mn concentration was varied between 0 and 0.40~mol\%, and a Cr-free Mn-doped series was synthesized as a reference. The concentrations were deliberately kept in the dilute regime, where a random dopant distribution would permit neither efficient energy transfer nor detectable same-ion pairing: even at the highest doping levels, the mean nearest-neighbor distance between randomly placed impurity ions is around 1.4~nm. Any interaction observed between the dopants at these dilutions therefore signals a non-random, associated arrangement.

\subsection{Structural analysis}

\begin{figure}[htbp]
        \includegraphics[width=\columnwidth]{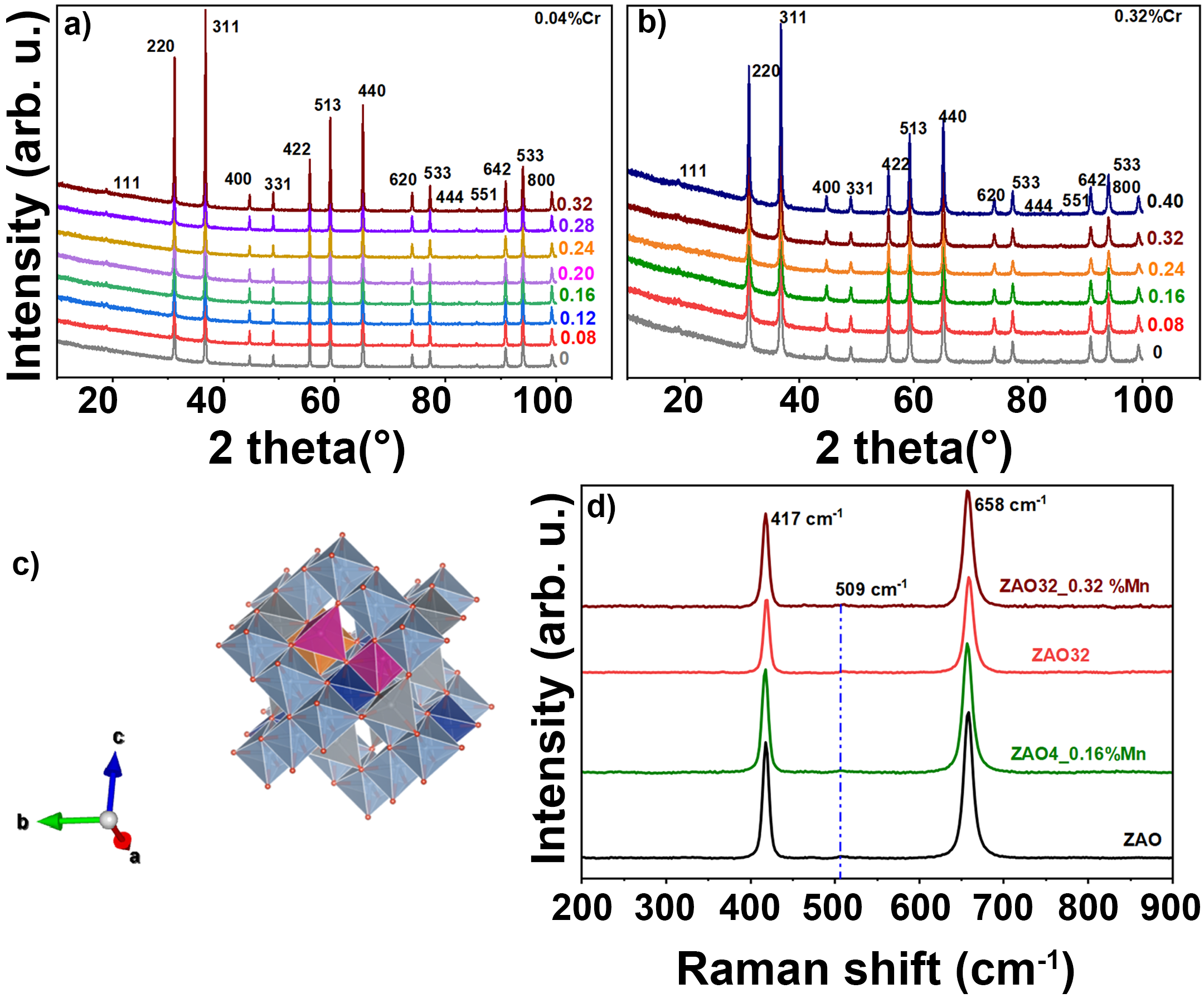}
        \caption{X-ray diffraction (XRD) patterns of \textbf{(a)} ZAO4 with $y$ mol\% Mn ($y$ = 0, 0.08, 0.12, 0.16, 0.20, 0.24, 0.28, and 0.32) and \textbf{(b)} ZAO32 with $y$ mol\% Mn ($y$ = 0, 0.08, 0.16, 0.24, 0.32, and 0.40) phosphors synthesized via a hydrothermal route. \textbf{(c)} Schematic representation of the crystal structure of \ch{ZnAl2O4} co-doped with \ch{Mn^2+} (purple), \ch{Mn^4+} (orange), and \ch{Cr^3+} (blue). \textbf{(d)} Representative Raman vibrational spectra of selected samples sintered at 1100~$^{\circ}$C.}
    \label{fig:xrd}
    \end{figure}

Powder X-ray diffraction and Raman microscopy both showed a well-crystallized system regardless of stoichiometry, with no detectable impurity phases or significant distortion. \textbf{Figures~\ref{fig:xrd}a and b} show the XRD patterns of single-phase zinc aluminate samples co-doped with various concentrations of Mn and Cr, with the corresponding Miller indices. All samples showed diffraction peaks indexed to the cubic spinel structure (space group $Fd\bar{3}m$, No. 227). \textbf{Figure~\ref{fig:xrd}c} shows the crystal structure. To analyze the unit-cell parameters and the crystallite sizes, Rietveld refinement was performed, and the results are presented in the Supplementary Information (\textbf{Table~\ref{tab:S1}}). Crystallite size and cell volume increased with [Mn]. This is attributed to the difference in ionic radii between \ch{Mn^2+} (0.83~\AA) and \ch{Zn^2+} (0.74~\AA), which causes lattice distortion and a gradual increase in the lattice constants.\cite{Trung2020,Tsai2013} In contrast, increasing Cr content showed no significant variation in lattice parameters but decreased crystallite size.

Raman spectroscopy probes lattice vibrations and can signal localized structural changes induced by doping in \ch{ZnAl2O4}.\cite{Beke2021} \textbf{Figures~\ref{fig:xrd}d and \ref{fig:S1}} present the Raman spectra of the two series. Two prominent bands\cite{Jain2020,DIppolito2015,Elhamdi2023} are observed at approximately 417~cm$^{-1}$ ($E_{g}$) and 657~cm$^{-1}$ ($A_{1g}/F_{2g}$). In between, a weak band at 509~cm$^{-1}$ is also present, ascribed to Zn--O stretching vibrations of $F_{2g}$ symmetry.\cite{MohacekGrosev2017,Fang2002}

    \begin{figure}[htbp]
        \includegraphics[width=\columnwidth]{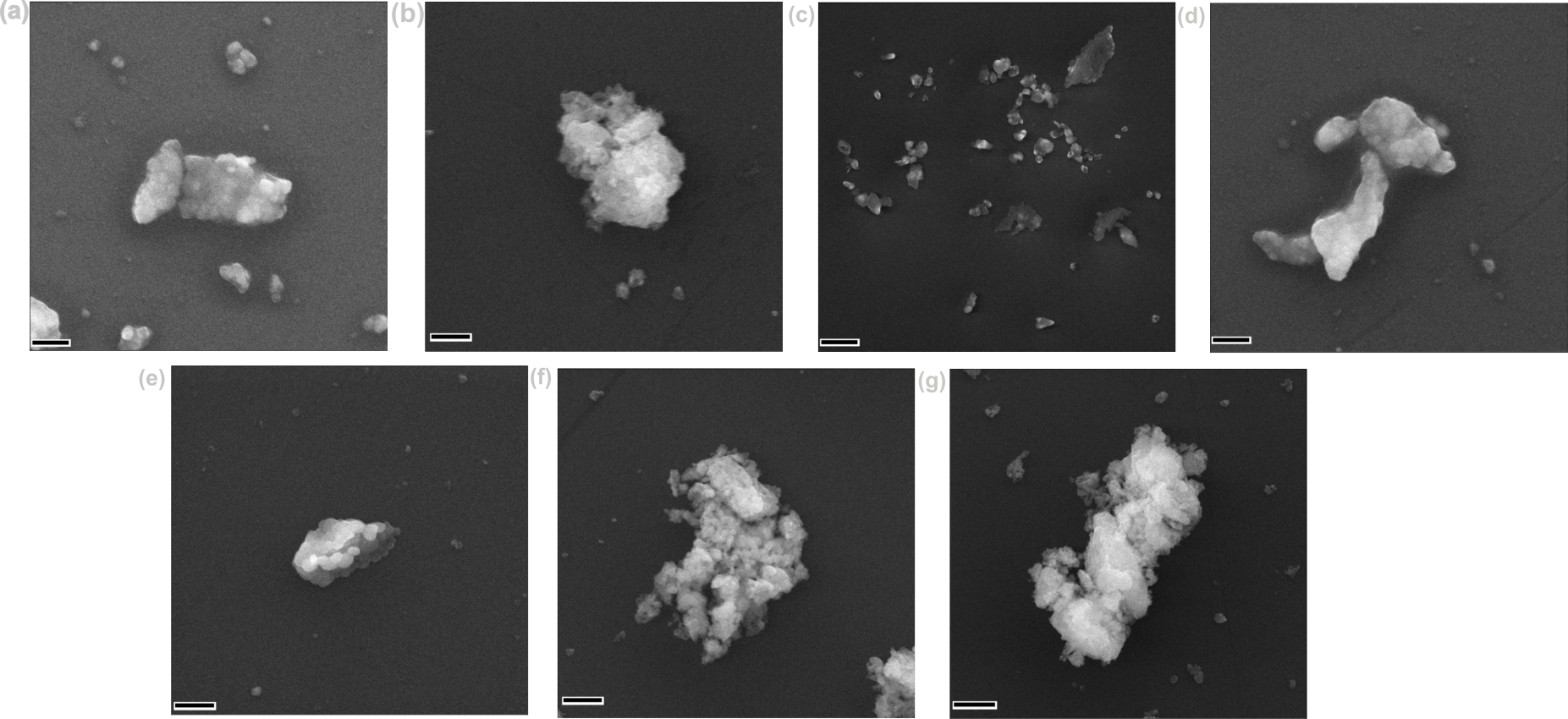}
        \caption{SEM micrographs of (a--f) ZAO4\_$y$\%Mn ($y$ = 0.08, 0.12, 0.16, 0.20, and 0.24) and (g) ZAO32\_$y$\%Mn ($y$ = 0.24 and 0.40). The scale bars represent 500~nm.}
    \label{fig:sem}
    \end{figure}

\begin{figure}[htbp]
        \includegraphics[width=\columnwidth]{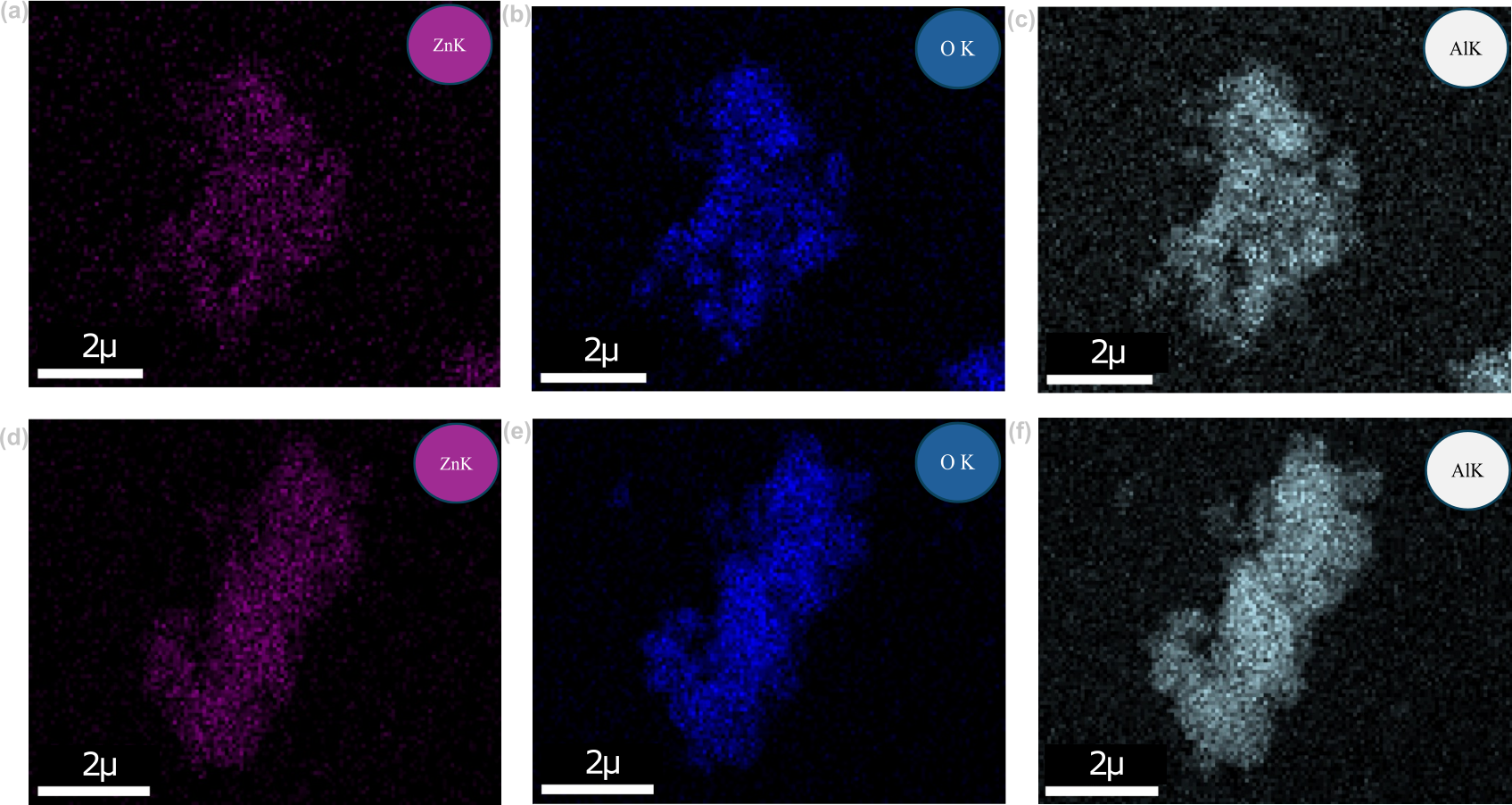}
        \caption{SEM-EDX elemental mapping of ZAO4\_0.24\%Mn (a--c) and ZAO32\_0.24\%Mn (d--f).}
        \label{fig:edx}
    \end{figure}

Scanning electron microscopy (SEM) was used to study particle size and morphology. \textbf{Figure~\ref{fig:sem}} depicts the SEM micrographs of selected ZAO4 and ZAO32 samples; the particle size distributions are summarized in \textbf{Figure~\ref{fig:S2}}. The particles exhibited a smooth spherical morphology and further agglomerated during the post-synthetic thermal treatment or during drop casting for the SEM studies. EDS and SEM-EDX revealed a very similar host stoichiometry across the samples (\textbf{Figure~\ref{fig:edx}}, Table~\ref{tab:S2}).

\subsection{Spectroscopic characterization}

\begin{figure*}[htbp]
        \includegraphics[width=\textwidth]{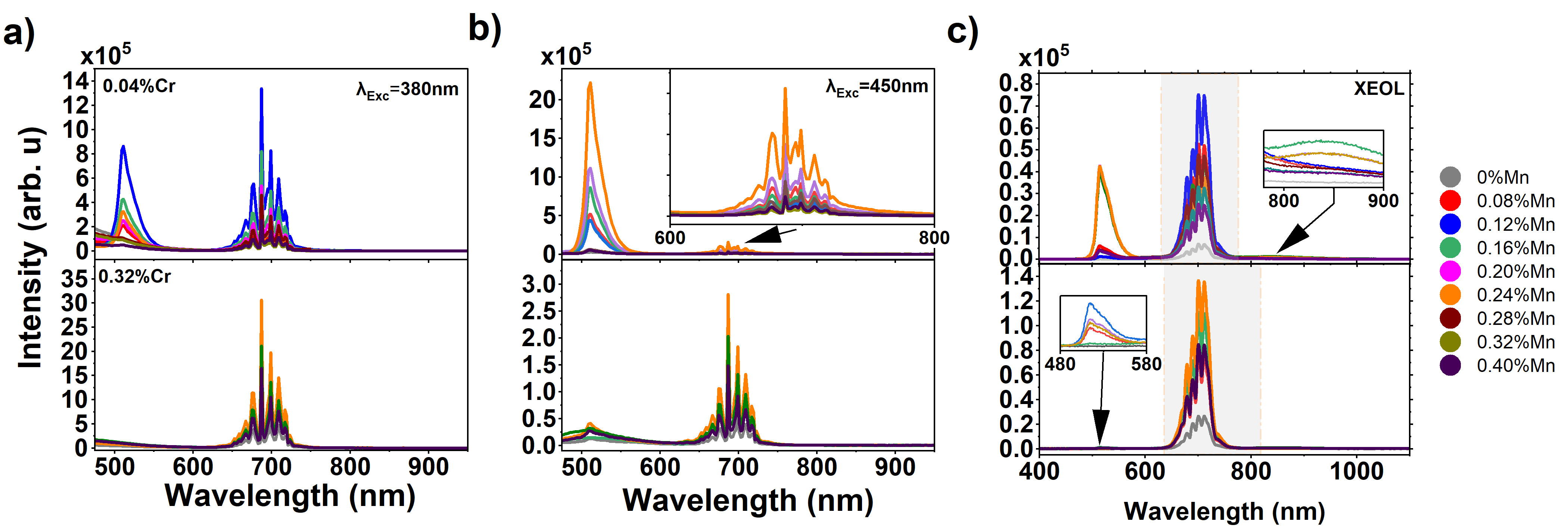}
        \caption{Photoluminescence spectra of the two series under (a) 380~nm and (b) 450~nm excitation and (c) X-ray excitation. Insets show the low-intensity emission bands.}
    \label{fig:pl}
    \end{figure*}

A series of spectroscopic techniques, including excitation--emission matrix (EEM) spectroscopy, PL, PLE, XEOL, and EPR, was employed to characterize how Mn ions affect the Cr emission intensity. The EEM allowed extraction of both the \ch{Mn^2+}-related emission peak around 510~nm and the \ch{Cr^3+}-related emission around 700~nm, along with their optimal excitation wavelengths (450~nm and 380~nm for Mn and Cr, respectively). The emission spectra of the two series under 380 and 450~nm excitation and under X-ray excitation are shown in \textbf{Figures~\ref{fig:pl}a--f}; the ZAO4 series also includes undoped ZAO, demonstrating the trace Cr impurities common in these spinels.\cite{Cornu2013,Lakshminarayana2011} The emission intensity of the ZAO4 series increased gradually and peaked at Mn concentrations of 0.12\% and 0.24\% under 380 and 450~nm excitation, respectively. To our surprise, no discernible \ch{Mn^2+} emission was detected in the ZAO32 samples under 380~nm excitation, and only weak green emission was observed under 450~nm excitation.

Under X-ray excitation, both series showed enhanced green and NIR emissions centered at 510~nm and 850~nm, respectively, alongside Cr signature peaks; however, the green and NIR emission intensities in ZAO32 samples were extremely weak, similar to PL. The data therefore suggest an interaction between the Mn and Cr dopants, as the increased Cr concentration significantly suppressed the Mn-related emissions, while the Cr-related PL peaks showed Mn concentration-dependent intensity variation despite the constant Cr content. The origin of the NIR emission in XEOL remains controversial,\cite{Xiao2025,VanBunningen2023,Liu2025} and is subject to further investigation.

\begin{figure}[htbp]
  \includegraphics[width=\columnwidth]{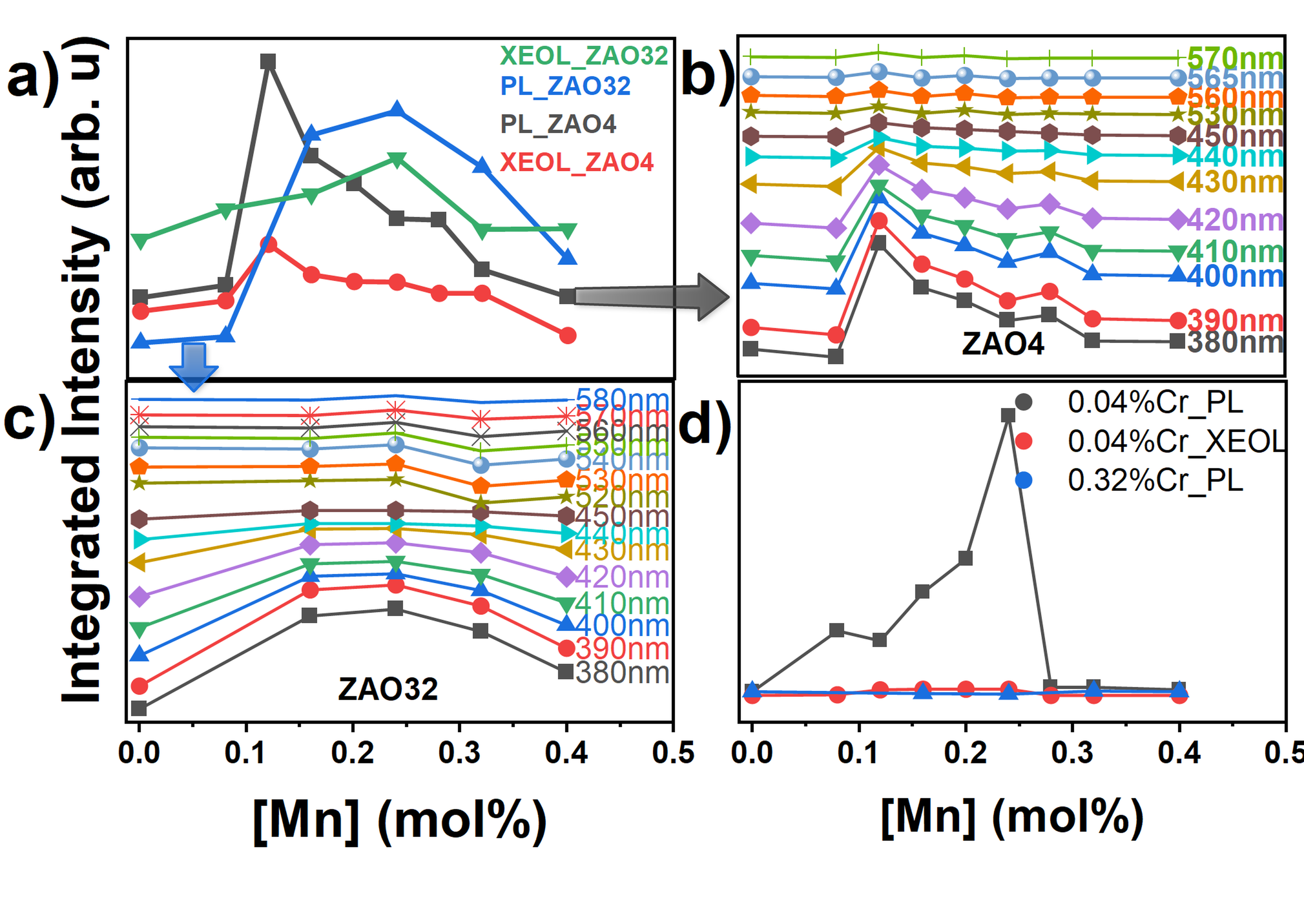}

        \caption{Integrated intensities of \ch{Cr^3+} as a function of [Mn], representing both PL and XEOL for the two sample sets (a); excitation-dependent intensities as a function of [Mn] in the range 300--570~nm for (b) ZAO4 and (c) ZAO32; (d) PL and XEOL integrated Mn emission intensities at 510~nm as a function of [Mn].}
    \label{fig:intensity}
    \end{figure}

\textbf{Figure~\ref{fig:intensity}} presents the variation in the emission intensities, obtained as the integrated areas of the corresponding bands (see Methods); the same analysis of the Mn-only reference series is shown in \textbf{Figure~\ref{fig:S3}}. The lowest \ch{Cr^3+} intensity was found in intrinsic ZAO, originating from trace Cr impurities in the precursors,\cite{Cornu2013,Khaidukov2020} while the highest intensities were observed at 0.12~mol\% and 0.24~mol\% Mn for ZAO4 and ZAO32, respectively. The relative intensity variation is more pronounced under UV excitation than under X-ray excitation. The Mn-only series reproduces both concentration trends. The \ch{Mn^2+} emission increased close to linearly with Mn content, peaked, and then fell at high concentration, while the trace \ch{Cr^3+} emission increased around 0.016~mol\% Mn and was suppressed by more than an order of magnitude above $\approx$0.06~mol\%. The enhancement therefore requires neither deliberate co-doping nor a particular Cr content, and it explains why Cr-related luminescence is routinely detected in nominally Cr-free \ch{ZnAl2O4} and other spinels. Because the chromium content of this series lies below the detection limit of atomic absorption spectrometry, gives no measurable EPR signal, and leaves the effective Mn range accordingly uncalibrated, the series is reported qualitatively and is not used for the quantitative analysis.

The \ch{Cr^3+}-related optical emission showed an excitation-wavelength dependence. As shown in \textbf{Figures~\ref{fig:intensity}b--c}, the observed [Mn] dependence of the Cr emission faded at 450~nm excitation and became almost constant, with a small increase where maxima were observed at lower excitation wavelengths.

Furthermore, the \ch{Cr^3+} emission spectra showed only marginal variation in shape, with relatively strong emission attributed to the zero-phonon line (ZPL) R and N2 lines.\cite{Singh2007}
This behavior contrasts with previous reports,\cite{Zhu2022} where Mn co-doping preferentially enhanced the intensities of the R1 and R3 lines while suppressing the N-lines and phonon sidebands (PSB, APSB). Nevertheless, the peak ratios of the R2/N2, R2/APSB, and N2/SPSB lines were calculated and are presented in \textbf{Figure~\ref{fig:S4}}. While R2/N2 is essentially constant, R2/APSB, which reflects the strength of the electron--phonon coupling between \ch{Cr^3+} and the lattice,\cite{Vink1998} spikes in all cases at the Mn concentration where the Cr emission intensity is at its maximum. Since electron--phonon coupling directly influences the photoluminescence intensity,\cite{Chen2018} the increase in the R2/APSB ratio, reflecting a transient weakening of the electron--phonon coupling, would offer a straightforward explanation for the sensitization effect. Beyond this coincidence of maxima at high excitation energies, however, the calculated ZPL/PSB ratios show no further correlation with the emission intensity; instead, they track the [Mn]-dependent Cr emission intensities recorded above 450~nm excitation.

\begin{figure}[htbp]
        \includegraphics[width=\columnwidth]{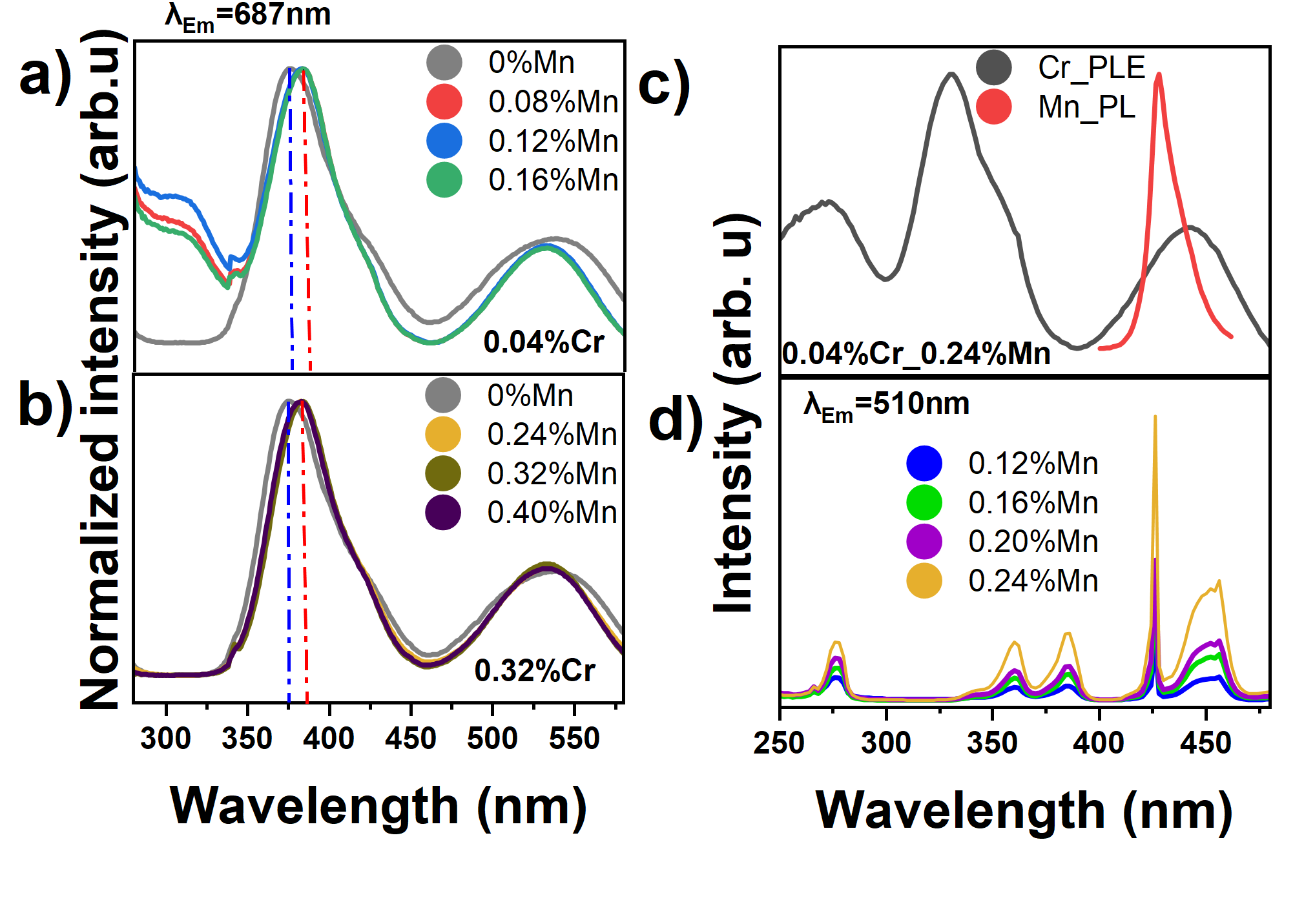}
        \caption{Photoluminescence excitation spectra of (a) ZAO4 and (b) ZAO32, both monitored at 687~nm. (c) PLE of ZAO4\_0.24\%Mn monitored at 687~nm and at the 510~nm Mn emission, showing the spectral overlap, and (d) PLE of Mn monitored at 510~nm.}
    \label{fig:ple}
    \end{figure}

\textbf{Figure~\ref{fig:ple}} shows the PLE spectra of the \ch{Mn^2+} and \ch{Cr^3+} co-doped phosphors. When monitoring the 687~nm emission of \ch{Cr^3+}, two prominent broad bands centered at 383~nm and 530~nm were observed and attributed to the d--d electronic transitions from $^{4}A_{2g}$ to $^{4}T_{1g}(^{4}F)$ and $^{4}T_{2g}(^{4}F)$, respectively.\cite{Si2021,Lorenzi2006} Upon Mn co-doping, a new absorption peak appeared at 310~nm, whose intensity increased with [\ch{Mn^2+}]. This peak was observed only in the ZAO4 samples. \textbf{Figure~\ref{fig:S5}} shows the emission spectra recorded under excitation between 300 and 340~nm. Both series exhibited spectra distinct from the Cr spectrum in ZAO, associated with \ch{Mn^4+} and octahedral \ch{Mn^2+}, respectively, rather than with Cr-related luminescence.\cite{Yang2023,Trung2020} Consequently, only \ch{Cr^3+} excitation bands that are not superimposed by tetrahedral \ch{Mn^2+} excitation peaks were detected in the 300--600~nm spectral range. A signature peak in the PLE spectra of \ch{Mn^2+} is a sharp peak at 427~nm (see \textbf{Figure~\ref{fig:ple}d} and \textbf{Figure~\ref{fig:S6}}), attributed to the $^{6}A_{1} \rightarrow {}^{4}A_{1}/{}^{4}E(^{4}G)$ transition. Although the PLE spectra showed a shoulder around 430~nm, it was also present in the Mn-free samples and is attributed to the trigonal ($D_{3d}$) splitting of the triply degenerate quartet terms.\cite{Mikenda1981} Si et al.\ reported similar findings in \ch{Mn^2+}/\ch{Cr^3+} co-doped \ch{ZnGa2O4}.\cite{Si2021} Other studies, however, have reported \ch{Mn^2+}-related PLE peaks alongside those of \ch{Cr^3+} when recording the \ch{Cr^3+} PLE,\cite{Zhu2022} which is direct evidence for energy transfer.

The most significant change in the \ch{Cr^3+} PLE spectra with [Mn], therefore, is the red shift of the $^{4}A_{2} \rightarrow {}^{4}T_{1}$ transition upon Mn doping. The shift was already detectable at the lowest [Mn] in the ZAO4 series and above 0.16\% Mn in the ZAO32 series. Notably, the shift was stepwise and showed no concentration dependence. The absolute and relative positions of the PLE peaks reflect the crystal field strength in these materials. The shift in $^{4}A_{2} \rightarrow {}^{4}T_{1}$ is due to a change in the interelectronic repulsion, represented by the Racah parameter $B$, which reflects the Cr--O bond covalency (nephelauxetic effect).\cite{Bai2020} \textbf{Table~\ref{tab:S3}} presents the crystal-field parameters $Dq/B$ and $B$, calculated with Equations~\eqref{eq:S1}--\eqref{eq:S5} from the band positions obtained by Gaussian decomposition of the PLE spectra (\textbf{Figures~\ref{fig:S7} and \ref{fig:S8}}). These values indicate that \ch{Cr^3+} is in a strong crystal field, $Dq/B > 2.07$.\cite{Song2024} With Mn introduced into the host lattice, the $Dq/B$ ratio shifted upward, indicating increased cationic ordering, while $B$ showed a stepwise drop, attributed to increased Cr--O bond covalency.

The emission peaks at 678~nm in the ZAO4 series and at around 650~nm in the ZAO32 series (\textbf{Figure~\ref{fig:S5}}) have been assigned to \ch{Mn^4+} in an octahedral environment\cite{Trung2020} and to \ch{Mn^2+} at the octahedral \ch{Al^3+} (antisite) position,\cite{Yang2023a} respectively. Both were prominent only at excitation wavelengths up to $\sim$340~nm, beyond which the characteristic \ch{Cr^3+} peaks emerge. The \ch{Mn^4+} emission is most intense at 0.12~mol\% Mn, whereas the octahedral \ch{Mn^2+} emission becomes pronounced only above 0.24\% Mn. Because the \ch{Mn^4+} emission overlaps with the \ch{Cr^3+} emission, its excitation bands contribute to the PLE spectra recorded at the \ch{Cr^3+} emission, accounting for the high-energy band observed in the ZAO4 series. Remarkably, two distinct and uncommon Mn occupations are thus stabilized in the two sample series.

Overall, the optical spectroscopy data show an apparent contradiction: the Cr emission intensity shows a clear [Mn] dependence, along with suppression of the \ch{Mn^2+} emission, and this dependence is confined to the Mn absorption window. Moreover, higher Cr concentrations require higher [Mn] to reach the maximum enhancement. Together, these observations support energy transfer from Mn to Cr ions. The PLE spectra of the Cr emission, however, exhibit no feature attributable to Mn absorption, nor is the emission selectively enhanced at any wavelength at which Mn absorbs. The presence of \ch{Mn^4+} and the antisite \ch{Mn^2+}, and their possible contribution to the Cr intensity variation, also require explanation.

\begin{figure}[htbp]
        \includegraphics[width=\columnwidth]{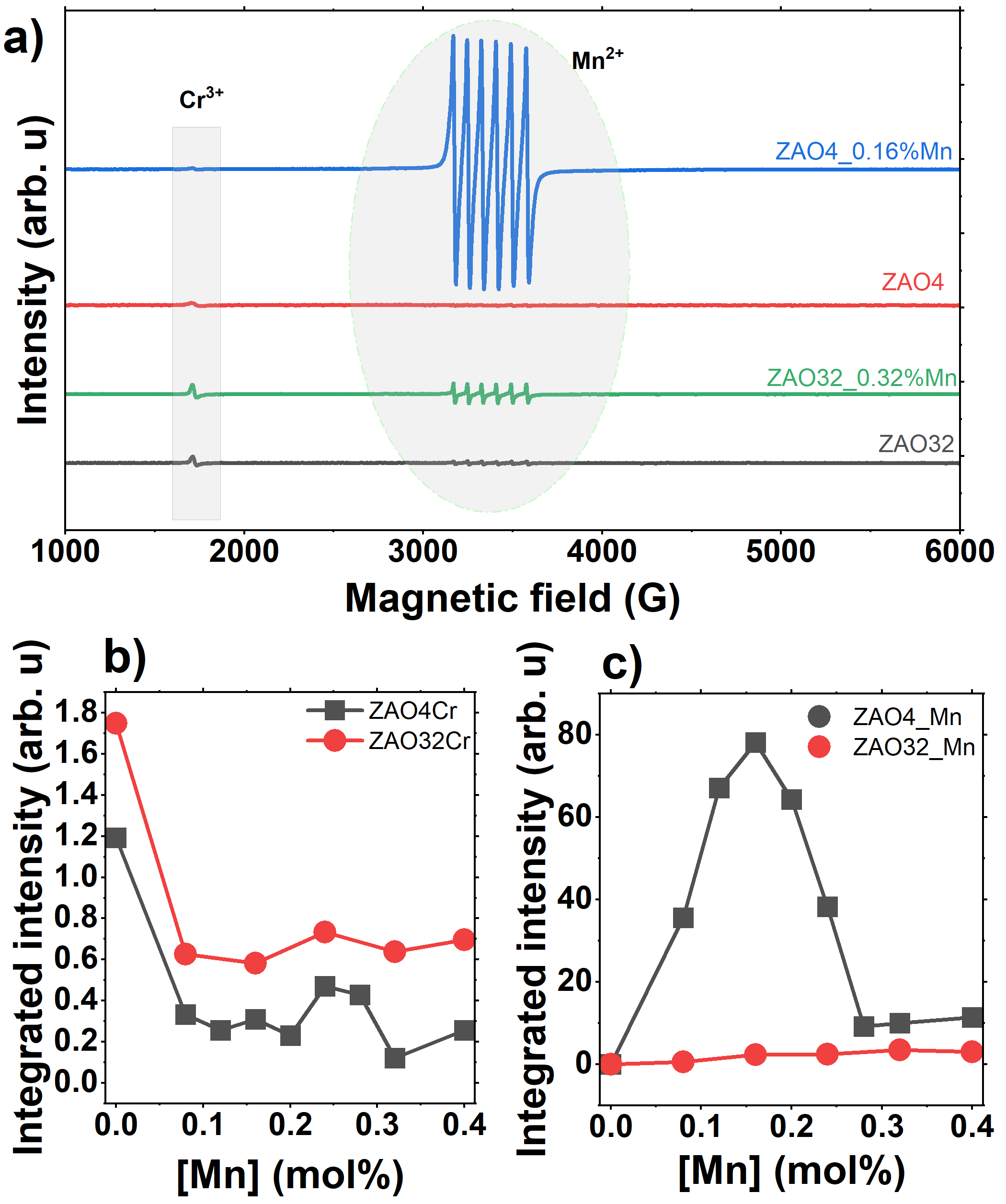}
        \caption{Room-temperature X-band EPR of the Mn/Cr co-doped \ch{ZnAl2O4} samples. (a) Representative spectra of the Mn-free ZAO4 and ZAO32 and the co-doped ZAO4\_0.16\%Mn and ZAO32\_0.32\%Mn samples between 100 and 600~mT; the shaded areas mark the low-field \ch{Cr^3+} line and the \ch{Mn^2+} hyperfine sextet. (b) Integrated intensity of the low-field \ch{Cr^3+} signal and (c) integrated intensity of the \ch{Mn^2+} sextet as a function of the Mn concentration for the ZAO4 and ZAO32 series, from spectra collected at 0.316~mW microwave power.}
    \label{fig:epr}
    \end{figure}

\textbf{Figure~\ref{fig:epr}} shows the room-temperature EPR spectra of the Cr-doped and Cr/Mn co-doped \ch{ZnAl2O4}. The spectrum of the Cr-doped sample (\textbf{Figure~\ref{fig:epr}a}) displayed the resonance features typical of a \ch{Cr^3+} (3d$^{3}$, $S = 3/2$) center in an axially distorted crystal field.\cite{Lorenzi2006,Ju2024} The co-doped samples also showed the characteristic six-line hyperfine sextet assigned to transitions within the $|{\pm}1/2\rangle$ Kramers doublet of \ch{Mn^2+} (3d$^{5}$, $S = 5/2$, $I = 5/2$), with \ch{Mn^2+} predominantly occupying the tetrahedral \ch{Zn^2+} sites.\cite{Cornu2014}

The most notable change in the Cr and Mn peaks was a decrease in intensity in the co-doped samples compared with the single-doped ones (\textbf{Figure~\ref{fig:epr}b,c}, \textbf{Table~\ref{tab:S4}}). The \ch{Cr^3+} low-field line of the co-doped ZAO4 samples retained about a quarter of its Mn-free value at every composition, with no systematic dependence on [Mn]. The Mn peak, however, reached a maximum in ZAO4 (0.16\% Mn) despite the increasing [Mn], and showed very low intensity in all ZAO32 samples, which reached a maximum at 0.32\% Mn. The integrated EPR intensity measures the number of resonant spins, so a decrease reflects a loss of the EPR-visible population. In the case of Mn, it can signal a change in the oxidation state, as \ch{Mn^3+} is EPR-silent; however, it does not explain the change in the \ch{Cr^3+} peak, which has no comparably stable alternative state in this host. The decreased intensities can also result from strong coupling to a neighboring paramagnetic ion, which redistributes the pair spins into states that are silent at X-band, as discussed below.

\section{Discussion}
\label{sec:discussion}

All applied methods suggest that the two dopants do not act as independent centers in ZAO under the studied conditions. We identified four features that might constrain the mechanism: (i) the stepwise, concentration-independent red shift of the \ch{Cr^3+} $^{4}A_{2} \rightarrow {}^{4}T_{1}$ excitation band (PLE, \textbf{Figure~\ref{fig:ple}a,b}), reaching the same final position in both series; (ii) the [Mn]-scaled optimum of the \ch{Cr^3+} emission enhancement and its confinement to $\lambda_{\mathrm{exc}} < 450$~nm (\textbf{Figure~\ref{fig:intensity}}); (iii) the variation of the EPR peak intensities across the series (\textbf{Figure~\ref{fig:epr}}); and (iv) the octahedral-Mn optical fingerprints (\textbf{Figure~\ref{fig:S5}}). We suggest that all four observations result from the association of octahedral Mn ions with \ch{Cr^3+}, and that the Cr--Mn associates are strongly exchange-coupled defect complexes whose population follows the law of mass action\cite{Kroger1956,Reiss1956} and whose Mn oxidation state is set by the available charge-compensating defects.

Because the argument rests on correlations across multiple spectroscopic data sets spanning the Mn and Cr concentrations, their statistical weight was assessed with nonparametric Spearman rank correlations,\cite{Bezerra2025} with significance levels derived from permutation tests, complemented by principal-component analysis of the standardized observable matrix. The principal-component analysis was consistent with the data set reducing to two independent factors --- an associate channel and a free-Mn channel. The results are reported in \textbf{Tables~\ref{tab:S5}--\ref{tab:S8}}, and the statistical analysis is described in detail in the \hyperref[sec:methods]{Materials and methods} section.

On its own, the deviation of an EPR signal from a linear concentration dependence, without the emergence of a detectable new line, points to exchange coupling: strong exchange redistributes the pair spins into total-spin multiplets whose transitions are spread over tens of thousands of gauss, which renders them silent at X-band. EPR therefore measures only the uncoupled spin populations. The spinel lattice, in turn, determines which Mn position can couple strongly to \ch{Cr^3+}. In the $Fd\bar{3}m$ structure, the \ch{AlO6} octahedra form an edge-sharing network:\cite{Sickafus1999} each octahedral cation has six octahedral neighbors at $a\sqrt{2}/4 \approx 2.86$~\AA, connected through a shared O--O edge that provides Mn--O--Cr bridges with direct cation--cation overlap, allowing strong magnetic coupling.\cite{Kanamori1959,Lee2002} The tetrahedral site, in contrast, couples to the octahedra only through a single $\sim$125$^{\circ}$ corner-sharing bridge at $\approx 3.35$~\AA, weaker by roughly an order of magnitude.\cite{Tomiyasu2004} The same inter-cation orbital overlap that mediates the superexchange also increases the covalency of the Cr--O--Mn bonds\cite{Anderson1959} and --- together with the symmetry lowering by the heterovalent neighbor and its compensating defect --- alters the effective Racah and crystal-field parameters that govern the absorption spectrum,\cite{Jorgensen1962} and with them, the PLE spectra. In other words, the same orbital configuration silences the EPR of the paired ions and shifts their absorption bands.

Indeed, the sharp low-field line of \ch{Cr^3+} retains about a quarter of its Mn-free intensity in every co-doped ZAO4 sample, so roughly three quarters of the chromium is paired. Since even the lowest Mn content is twice the Cr content in ZAO4, the associate population is limited by chromium and is already the majority at the lowest [Mn].
At the same time, the weak variation of the R/N line ratio of the Cr PL shows that the emission is not suppressed for the Cr ions that retain an EPR signal.

On the Mn side, the associated Mn in ZAO4 amounts to the order of the Cr content itself. The free \ch{Mn^2+} signal therefore continues to grow with [Mn] and peaks at 0.16\% Mn (the origin of this maximum is discussed below). In ZAO32, the sextet stays at 1--2\% of the ZAO4 level --- most of the Mn is in EPR-silent configurations --- yet follows the same pattern, increasing to a maximum near 0.28--0.32\% Mn. The absence of any Mn structure other than the ideal tetrahedral site in the EPR spectra, while octahedral \ch{Mn^2+} and \ch{Mn^4+} are detected in PL, also supports Mn residing in EPR-silent configurations.

Within the associate, $^{4}A_{2} \rightarrow {}^{4}T_{1}$ moves down while $^{4}A_{2} \rightarrow {}^{4}T_{2}$ moves slightly up: $Dq$ increases marginally while $B$ of the absorbing ensemble falls by 10--17\%, the nephelauxetic consequence of the higher covalency around a heterovalent cation sharing an octahedral edge.\cite{Jorgensen1962} \ch{Cr^3+} occupies a centrosymmetric site whose d--d absorption is electric-dipole forbidden and only vibronically allowed;\cite{Sugano1970} the adjacent octahedral Mn and its compensating defect remove the inversion symmetry and admix odd-parity charge-transfer character into the d--d states.\cite{Judd1962,Ofelt1962} Consistent with this, the ZPL/PSB ratio slightly increases due to the larger Debye--Waller factor, i.e., a smaller effective Huang--Rhys parameter. Because the admixture weakens with increasing energy separation from the odd-parity states,\cite{Judd1962} the absorption gains strength strongly for $^{4}A_{2} \rightarrow {}^{4}T_{1}$ ($\approx$26,000 cm$^{-1}$, 385~nm), weakly for $^{4}A_{2} \rightarrow {}^{4}T_{2}$ ($\approx$18,800 cm$^{-1}$, 532~nm), and not at all for direct $^{2}E$ excitation. This reproduces the observed intensity increase and its excitation dependence as well as the marginal changes in the emission spectra ($^{2}E \rightarrow {}^{4}A_{2}$ transition).

Exchange coupling also relaxes the single-ion spin selection rules and mixes the excited states of the two ions into common pair states;\cite{McCarthy1988,Ferguson1966,Tanabe1965} a strongly coupled associate then behaves as a single emitter that relaxes internally and emits from its lowest excited state, with a shorter lifetime.\cite{Zhu2021,Chen2023,Su2018} That state is the Cr-derived $^{2}E$ at $\approx$14,550 cm$^{-1}$, below every Mn-related level: the tetrahedral $^{4}T_{1}$ lies at $\approx$19,600 cm$^{-1}$, while the octahedral \ch{Mn^2+} $^{4}T_{1}$ at $\approx$15,400 cm$^{-1}$ and the \ch{Mn^4+} $^{2}E$ at $\approx$14,750 cm$^{-1}$ are only about $4\,k_{\mathrm{B}}T$ higher, making the 678 and 687~nm channels nearly thermalized relaxation paths. This also explains why the Cr PLE shows no Mn-level excitation peaks despite the Mn-induced increase of the Cr PL intensity: the absorption of the associate is dominated by its allowed, Cr-derived and charge-transfer transitions; the Mn it contains is octahedral, whose transitions overlap with those of Cr, while transitions of predominantly Mn character remain spin-forbidden even in the pair.

Association alone, however, does not explain the presence of a maximum, more specifically the decrease in the PL and EPR intensities as a function of [Mn]. The coincidence of the two maxima points to a second mechanism: a collapse of the bright configuration at a given Mn content, or [Mn]/[Cr] ratio. Although each Cr offers six octahedral neighbor sites, occupation is expected to stop at one Mn per Cr, because every octahedral Mn --- divalent or tetravalent --- requires charge compensation. The oxidation state of the associated Mn is then set by the compensating defects available to complete the complex. \ch{Mn^4+} on the \ch{Al^3+} site is a donor-type substitution requiring acceptor compensation by cation vacancies or Zn-on-Al antisites, whereas oxygen vacancies favor the reduced species.\cite{Du2014,Brik2013,Huang2020} Since the negatively charged antisite binds preferentially\cite{DeVos2016} to \ch{Cr^3+}, the number of acceptor states available around the Cr centers is large in ZAO4, allowing oxidation of the associated Mn to \ch{Mn^4+}. In ZAO32 the same acceptor content is spread over an eightfold larger Cr population, so most Cr--Mn units cannot recruit a dedicated acceptor and the associated Mn remains divalent.
As both Mn species occupy the same lattice position, their perturbation of the neighboring \ch{Cr^3+} is essentially identical. We speculate that the collapse results from a change in the charge-compensation equilibrium of the crystal: above a threshold Mn content, the available compensation re-partitions most of the Mn population into EPR-silent octahedral configurations, which darkens the associates and quenches the emissions. The populations of the charged Mn configurations depend exponentially on the Fermi-level position,\cite{VandeWalle2004,Freysoldt2014} while the balance between the neutral configurations --- tetrahedral \ch{Mn^2+} and octahedral \ch{Mn^3+} --- follows the cation chemical potentials fixed by the same defect network; a modest shift of this coupled equilibrium therefore redirects Mn between sites and valences. Above a Cr-dependent threshold, the Mn population re-partitions from \ch{Mn^2+} into EPR-silent octahedral configurations, most plausibly into non-Kramers \ch{Mn^3+}, which is isovalent on the \ch{Al^3+} site and requires no external compensation, and/or self-compensated octahedral pairs, \ch{Mn^2+}--O--\ch{Cr^3+}--O--\ch{Mn^4+}. Because Cr-bound defects participate in the compensation balance, a higher Cr content shifts the threshold to a higher Mn content. The $\approx$850~nm emission observed under X-ray excitation in ZAO4 may indicate such \ch{Mn^3+} population. \ch{Mn^3+} in spinels and related oxides shows a weak room-temperature emission in this spectral region,\cite{Kazarinov2025} although whether the \ch{Mn^3+} pre-exists or is created by the ionizing radiation cannot be decided at this stage of the research.

\begin{figure}[htbp]
        \includegraphics[width=\columnwidth]{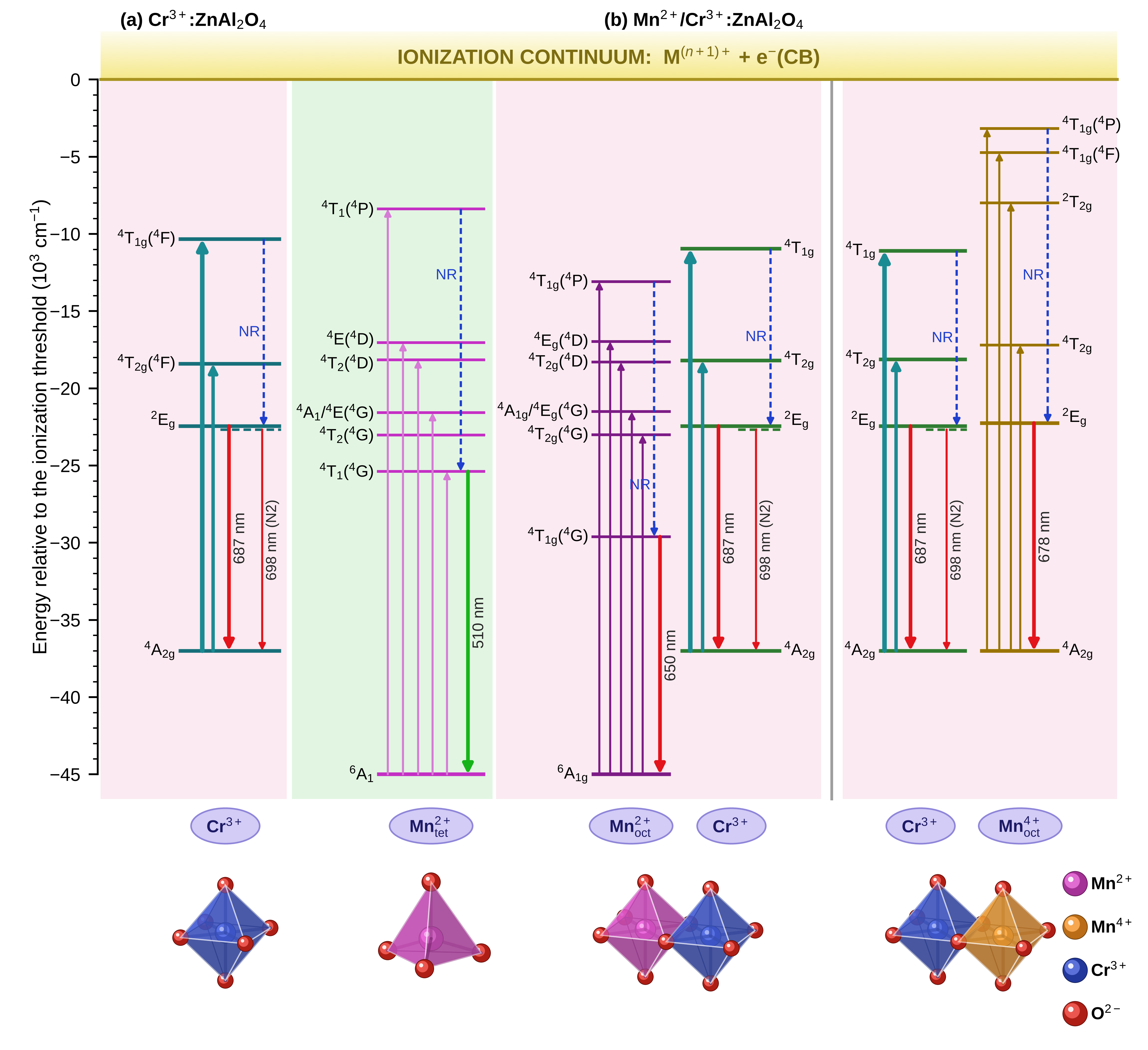}
        \caption{Schematic energy-level diagram of (a) \ch{Mn^2+} and \ch{Cr^3+} in \ch{ZnAl2O4} and (b) tetrahedral \ch{Mn^2+} and the octahedral \ch{Mn^2+}--\ch{Cr^3+} and \ch{Cr^3+}--\ch{Mn^4+} complexes in Mn/Cr co-doped \ch{ZnAl2O4}. Energies are given relative to the ionization threshold of each center, $\mathrm{M}^{(n+1)+} + \mathrm{e}^{-}(\mathrm{CB})$. Solid upward arrows denote excitation, dashed arrows non-radiative (NR) relaxation, and solid downward arrows emission, labeled with the emission wavelength; the dashed level just below $^{2}E_{g}$ belongs to the N2 site of \ch{Cr^3+}. Level energies were derived from the PLE spectra (Figure~\ref{fig:ple}, Figure~\ref{fig:S6} and Table~\ref{tab:S3}); the higher \ch{Mn^4+} levels were taken from the literature. The coordination polyhedra of the centers are shown below the diagram.}
    \label{fig:levels}
    \end{figure}

To summarize the excitation and relaxation pathways, the relevant electronic transitions are shown schematically in \textbf{Figure~\ref{fig:levels}}. The energy levels were derived from the PLE spectra (Figure~\ref{fig:ple} and Figure~\ref{fig:S6}). For \ch{Cr^3+}, the spin-allowed $^{4}A_{2} \rightarrow {}^{4}T_{2}$ and $^{4}A_{2} \rightarrow {}^{4}T_{1}$ transitions were taken from the PLE spectra, together with the transition from the \ch{Cr^3+} ground state to the conduction band at approximately 37,000 cm$^{-1}$. The charge-transfer band of tetrahedral \ch{Mn^2+} lies at about 45,000 cm$^{-1}$.\cite{Yin2026} Octahedral \ch{Mn^2+} follows the same scheme as its tetrahedral counterpart, except that its emitting $^{4}T_{1}$ level lies lower because of the stronger octahedral crystal field. For octahedral \ch{Mn^4+}, part of the level structure could be extracted from the PLE spectrum monitored at 678~nm. Its excitation bands, however, overlap substantially with the \ch{Cr^3+} absorption bands; therefore, the remaining levels were taken from the literature.\cite{Gao2021,VanQuang2026}

\section{Conclusions}

The luminescent response of Cr is influenced by its coexistence with Mn in the spinel host lattice. We propose that, beyond energy transfer, defect-complex formation influences the optical properties of both dopants, increasing Cr absorbance and reducing recombination through Mn channels. The defect complex consists of an octahedral Mn paired with Cr, and because octahedral Mn needs charge compensation to preserve overall neutrality, its oxidation state depends on the absolute Mn concentration and the relative Mn:Cr ratio.

The stage of synthesis at which this equilibrium is established remains unclear, as it cannot be determined from the available data alone. The co-precipitation of Zn/Al at pH 8--10 proceeds through a Zn--Al layered double hydroxide (LDH).\cite{Chen2005,Chen2002,Ishii2016,Zhao2012} LDH sheets are all-octahedral and edge-sharing with double hydroxo bridges, and the \ch{M^2+}/\ch{M^3+} cations order into a honeycomb in which every trivalent ion is surrounded by six divalent neighbors.\cite{Radhakrishnan2021,Dionigi2016,Carrasco2020} Crystalline Mn(II)Cr(III)-LDH itself forms by co-precipitation\cite{Timar2017} at pH 10, heterometallic M(II)--(OH)--Cr(III) condensation is favored over the homometallic route,\cite{Packter1985,Oestreicher2019} and the slow ligand exchange of Cr(III) stabilizes a bridge once it has formed.\cite{Casey2003} A Mn ion beside Cr in such a sheet therefore already possesses the double hydroxo bridge of the spinel associate, and the topotactic LDH-to-spinel transformation delivers it to the octahedral sublattice.\cite{Rabe2021,Ortega2017} This would also explain why hydrothermally synthesized spinels differ from their counterparts prepared via the solid-state route. The 1100~$^{\circ}$C anneal, however, allows cation rearrangement: the diffusion lengths extrapolated from the closest measured systems are about 0.1--1~\textmu m for hour-scale annealing.\cite{Weeks1978,Weber1977,Kashii1999} Therefore, the associated population can also equilibrate during annealing.

The association in solution and subsequent precipitation, however, offer a concentration-dependent explanation for the drop in Mn EPR and Cr PL intensities once a certain concentration is reached. Mn(II) oxidation at alkaline pH is negligible homogeneously but proceeds heterogeneously and autocatalytically on the Mn(III/IV) (hydr)oxides it produces,\cite{Tekerlekopoulou2013,Kessick1975} and concentrated Mn(II) solutions under such conditions yield the Mn(III)-bearing phases MnOOH and hausmannite \ch{Mn3O4}.\cite{Noda2019,Wang2015} On the other hand, annealing also eliminates many defects in the system and, with them, their charge compensation, which is necessary to stabilize \ch{Mn^2+} and \ch{Mn^4+} in the octahedral position.

This behavior in the nanostructures --- two dilute species, an associated form, and concentrations that respond to each other --- is the solid-state analog of a complexation equilibrium and follows the law of mass action:\cite{Kroger1956,Reiss1956} for $\mathrm{Mn_{f}} + \mathrm{Cr_{f}} \rightleftharpoons \{\mathrm{Mn\text{--}Cr}\}$, the equilibrium concentrations obey
\begin{equation}
K_{\mathrm{d}} = \frac{[\mathrm{Mn_{f}}][\mathrm{Cr_{f}}]}{[\{\mathrm{Mn\text{--}Cr}\}]},
\label{eq:Kd}
\end{equation}
where the subscript f denotes the free, unassociated species and all concentrations are expressed as nominal cation fractions. The EPR spin counts give $K_{\mathrm{d}} \approx (2\text{--}5) \times 10^{-4}$ across 0.08--0.20\% Mn, i.e., an association free energy $\Delta G_{\mathrm{a}} = k_{\mathrm{B}}T \ln(1/K_{\mathrm{d}}) \approx 8\,k_{\mathrm{B}}T$ at the temperature at which the equilibrium was last established --- $\approx$0.6--0.7~eV for freezing near 900--1000~K, or $\approx$0.35~eV if the equilibrium closed at the hydrothermal temperature (230~$^{\circ}$C, 503~K). The value of $\Delta G_{\mathrm{a}}$ lies in the range of independently measured Mn--Cr affinities: the formation of \ch{MnCr2O4} from MnO and \ch{Cr2O3} is favored by 44~kJ/mol (0.46~eV) at 1573~K,\cite{Timucin1992} \ch{Mn^2+} prefers a chromite over an aluminate framework by 19--25~kJ/mol (0.20--0.25~eV),\cite{Timucin1992,Petric2007} and computed dopant-pair binding energies in related compounds are in the 0.4--0.6~eV range;\cite{Sousa2024,DeVos2016} $\Delta G_{\mathrm{a}}$ therefore reads as the chemical Mn--Cr affinity plus the electrostatic binding of the compensating defect.

Despite the unambiguous evidence of the solution phase defect formation, co-doping is a versatile tool for defect engineering. It allows precise tuning of material properties, including the optical response to different stimuli. It requires, however, the understanding and controlled manipulation of the relative position of and interaction between the dopants. Hydrothermal synthesis of materials or precursors might allow further tuning of these properties through interactions among components in solution, which do not exist in the solid state.

\section*{Acknowledgements}
This work was supported by the Doctoral Excellence Fellowship Programme (DCEP), funded by the National Research, Development and Innovation Fund of the Ministry of Culture and Innovation and by the Budapest University of Technology and Economics. The Hungarian research infrastructure was provided by the Hungarian Academy of Sciences. The research was funded by project no.~TKP2021-NVA-04 of the Ministry of Innovation and Technology of Hungary, financed from the National Research, Development and Innovation Fund (NKFIH) under the TKP2021-NVA funding scheme, and by the Quantum Information National Laboratory sponsored by the Ministry of Innovation and Technology of Hungary via the NKFIH. This work was also supported by the National Research, Development and Innovation Office of Hungary (NKFIH), and by the Ministry of Culture and Innovation (grant no.~149457).

\section{Experimental section}
\label{sec:experimental}

\subsection{Materials and methods}
\label{sec:methods}

\ch{Zn(NO3)2 * 6 H2O} (reagent grade, 98\%), \ch{Al(NO3)3 * 9 H2O} (trace-metal basis), \ch{Cr(NO3)3 * 9 H2O} (trace-metal basis, 99.99\%) and \ch{Mn(NO3)2 * $x$ H2O} (trace-metal basis, 99.99\%) from Sigma-Aldrich, HCl (37\%, VWR AnalaR NORMAPUR), \ch{NH3} (32\%, VWR HiPerSolv CHROMANORM), high-purity 18~M$\Omega\cdot$cm Millipore type-1 water (DI water), and 2-propanol (IPA, 99.8\%, VWR HiPerSolv CHROMANORM), used to precipitate the formed particles, were the starting materials. All reagents were used without further purification.

Stock solutions of 2~M \ch{Al(NO3)3}, 1~M \ch{Zn(NO3)2}, 1~mM \ch{Cr(NO3)3}, and 1~mM \ch{Mn(NO3)2} were prepared. To synthesize ZAO\_$x$Cr\_$y$Mn ($x$ = 0.04 or 0.32~mol\% Cr; $y$ = 0--0.40~mol\% Mn; percentages are nominal cation fractions relative to Al and Zn, respectively), stoichiometric volumes of the Zn and Al stock solutions were measured into cleaned PTFE vessels and stirred for 30~min; the Cr and Mn stock solutions were then added sequentially in the required proportions and stirred for a further 30~min. Ammonia solution (32\%) was added to adjust the pH to 9, and the volume was adjusted to 15~mL with deionized water. The sealed autoclaves were held at 230~$^{\circ}$C for 13~h (Berghof DAH heating block, BTC-3000 temperature controller). The white precipitate was collected by centrifugation (8000~rpm, 10~min), washed three times with IPA containing 0.01~M HCl and three times with IPA, dried at 60~$^{\circ}$C, ground, and sintered in air at 1100~$^{\circ}$C for 6~h in a muffle furnace.

\paragraph{XRD} Patterns were recorded with a Huber G670 Guinier Imaging Plate Camera using Cu K$\alpha_{1}$ radiation ($\lambda$ = 1.54056~\AA) from a focusing Johansson monochromator. Glass capillaries of diameter 0.5~mm and wall thickness of 0.01~mm were used to hold the samples. Further analysis of crystallite size was performed by Rietveld refinement using the Topas Academic program.

\paragraph{SEM} A TESCAN MIRA3 electron microscope equipped with an Element EDS system was used to analyze the morphology and extract the elemental composition and distribution.

\paragraph{Raman spectroscopy} Spectra were recorded on powdered samples on a Renishaw InVia confocal Raman microscope with 532~nm laser excitation and a diffraction grating of 1800 lines per millimeter.

\paragraph{Electron paramagnetic resonance} Measurements were conducted on a Magnettech miniscope MS-400 spectrometer equipped with a TE102 rectangular resonator. Spectra were recorded with a field sweep from 100 to 600~mT at the X-band frequency of 9.39~GHz, a microwave power of 0.1~mW, and a modulation amplitude of 0.1~mT. Power-dependent spectra were recorded between 0.05 and 31.6~mW, and the integrated intensities were referenced to a DPPH standard. The EasySpin MATLAB package was used to simulate the spectra.

\paragraph{Photoluminescence} Photoluminescence measurements were performed on solid powders placed on a sample holder with a diameter of 8.5~mm. Samples of equal mass were evenly spread on the holder and placed on a stage, where they were irradiated with 365~nm (excitation) UV light from a THORLABS DC40 LED driver. An optical system collects the luminescence, which is then passed through long-pass filters to cut off the excitation and transmitted to the Ocean Optics QEpro spectrometer through a QP600-025-UV-BX optical fiber (NA 0.22). EEM and PLE spectra were collected with a Horiba Jobin Yvon spectrofluorometer. The PL and XEOL spectra of Mn and Cr were integrated to calculate their intensities.

\paragraph{XEOL} XEOL was performed with non-monochromatized unfiltered X-ray radiation from a sealed X-ray tube of Mo anode driven at 50~kV and 20~mA. An X-ray beam of 7.5~mm diameter was used to irradiate the sample at an inclination angle of 45$^{\circ}$ resulting in an approximately 60~mm$^{2}$ irradiated area.

\paragraph{Statistical analysis} The co-variation of the spectroscopic data with the Mn content was quantified with nonparametric statistics, chosen because the concentration series are short ($n = 5$--9) and the expected relations are monotonic rather than linear. For each pair of experimental outputs, the Spearman rank correlation coefficient $r_{\mathrm{s}}$ was computed, and its two-sided significance was estimated by a Monte Carlo permutation test. One variable was randomly permuted $2 \times 10^{4}$ times, and the p-value was taken as the fraction of permutations yielding $|r_{\mathrm{s}}|$ at least as large as observed. This procedure remains valid for sample sizes at which parametric correlation tests are unreliable. The outputs were the integrated intensities of the free tetrahedral \ch{Mn^2+} EPR sextet (an inverse-variance-weighted mean over the three lowest microwave powers, 0.10--1.0~mW, normalized per unit sample mass) and of the low-field \ch{Cr^3+} EPR line intensity, the \ch{Mn^2+} and \ch{Cr^3+} photoluminescence, XEOL, and the 678~nm \ch{Mn^4+} emission for the ZAO4 series, and the \ch{Cr^3+} photoluminescence, XEOL and 650~nm octahedral-\ch{Mn^2+} channels for the ZAO32 series. Because nineteen pairwise correlations are reported in Tables~\ref{tab:S5} and \ref{tab:S6}, their p-values are given unadjusted and interpreted descriptively; Holm and Benjamini--Hochberg adjustments over the full set are quoted in the table captions, and only the \ch{Cr^3+} PL--XEOL pair remains individually significant after either correction. The association model as a whole was instead tested once per series: the statistic was the mean $r_{\mathrm{s}}$ over the pairs the model predicts to co-vary positively (\ch{Cr^3+} PL--XEOL, \ch{Cr^3+} PL--\ch{Mn^4+}, XEOL--\ch{Mn^4+} and \ch{Mn^2+} EPR--510~nm \ch{Mn^2+} PL for ZAO4; \ch{Cr^3+} PL--XEOL and 650~nm--[Mn] for ZAO32), and its one-sided p-value was obtained by permuting each dataset independently ($2 \times 10^{4}$ iterations). Spearman partial correlations controlling for Mn content were computed by rank-transforming the variables, regressing out the ranked concentration, and correlating the residuals, with significance assessed by permutation of the residuals. Principal-component analysis of the standardized (z-scored) data matrix of the ZAO4 series (six compositions, $y \leq 0.24\%$; six observables) is reported as an exploratory summary --- six observations cannot determine six-variable loadings precisely --- and its stability was checked by leave-one-out resampling. The correlation coefficients, permutation p-values, principal-component elements, partial correlations and pattern tests are collected in \textbf{Tables~\ref{tab:S5}--\ref{tab:S8}}.

\bibliographystyle{unsrtnat}
\bibliography{references}

\clearpage
\setcounter{figure}{0}
\renewcommand{\thefigure}{S\arabic{figure}}
\setcounter{table}{0}
\renewcommand{\thetable}{S\arabic{table}}
\setcounter{equation}{0}
\renewcommand{\theequation}{S\arabic{equation}}
\setcounter{section}{0}
\renewcommand{\thesection}{S\arabic{section}}

\section*{Supplementary Information}

\begin{table}[htbp]
\centering
\caption{Rietveld refined parameters: lattice constant ($a=b=c$), cell volume ($V$), crystallite size ($d$) and the weighted profile R-factor ($R_{\mathrm{wp}}$) of Mn-Cr co-doped zinc aluminate annealed at 1100~$^{\circ}$C.}
\label{tab:S1}
\small
\begin{tabular}{@{}lcccc@{}}
\toprule
{[}Mn{]} & $a$ & $V$ & $d$ & $R_{\mathrm{wp}}$ \\
(mol\%) & (\AA) & (\AA$^{3}$) & (nm) & (\%) \\
\midrule
\multicolumn{5}{l}{\textit{$x$ = 0.04\% Cr}} \\
0 & 8.087 & 528.927 & 88.7 & 4.10 \\
0.08 & 8.089 & 529.360 & 116.7 & 4.46 \\
0.12 & 8.090 & 529.531 & 117.6 & 4.26 \\
0.16 & 8.089 & 529.328 & 116.3 & 4.40 \\
0.20 & 8.088 & 529.215 & 102.6 & 4.20 \\
0.24 & 8.089 & 529.276 & 102.2 & 4.12 \\
0.28 & 8.089 & 529.341 & 113.9 & 4.50 \\
0.32 & 8.088 & 529.162 & 102.8 & 4.40 \\
\multicolumn{5}{l}{\textit{$x$ = 0.32\% Cr}} \\
0 & 8.080 & 527.488 & 38.0 & 4.24 \\
0.08 & 8.080 & 527.442 & 39.0 & 4.29 \\
0.16 & 8.082 & 527.875 & 41.9 & 4.36 \\
0.24 & 8.083 & 528.052 & 40.9 & 4.48 \\
0.32 & 8.083 & 528.026 & 40.3 & 4.36 \\
0.40 & 8.079 & 527.319 & 39.1 & 4.49 \\
\bottomrule
\end{tabular}
\end{table}

\begin{table}[htbp]
\centering
\caption{SEM-EDS elemental composition of selected samples of ZAO\_0.04\%Cr\_$y$\%Mn at different annealing temperatures.}
\label{tab:S2}
\small
\begin{tabular}{lcccccc}
\toprule
Sample & Zn at\% & Al at\% & O at\% & \%Total & Error\% Mn & Error\% Cr \\
\midrule
0.08\%Mn & 13.24 & 26.13 & 60.63 & 100 & 65.79 & 62.61 \\
0.16\%Mn & 13.51 & 26.85 & 59.64 & 100 & 58.49 & 59.55 \\
0.20\%Mn & 14.38 & 28.15 & 59.47 & 100 & 57.82 & 59.01 \\
0.24\%Mn & 12.56 & 24.88 & 62.41 & 100 & 60.42 & 61.02 \\
0.32\%Mn & 12.30 & 24.79 & 64.91 & 100 & 89.42 & 65.40 \\
0.40\%Mn & 11.15 & 22.28 & 66.36 & 100 & 61.65 & 57.91 \\
\bottomrule
\end{tabular}
\end{table}

\begin{figure}[htbp]
\centering
\includegraphics[width=\columnwidth]{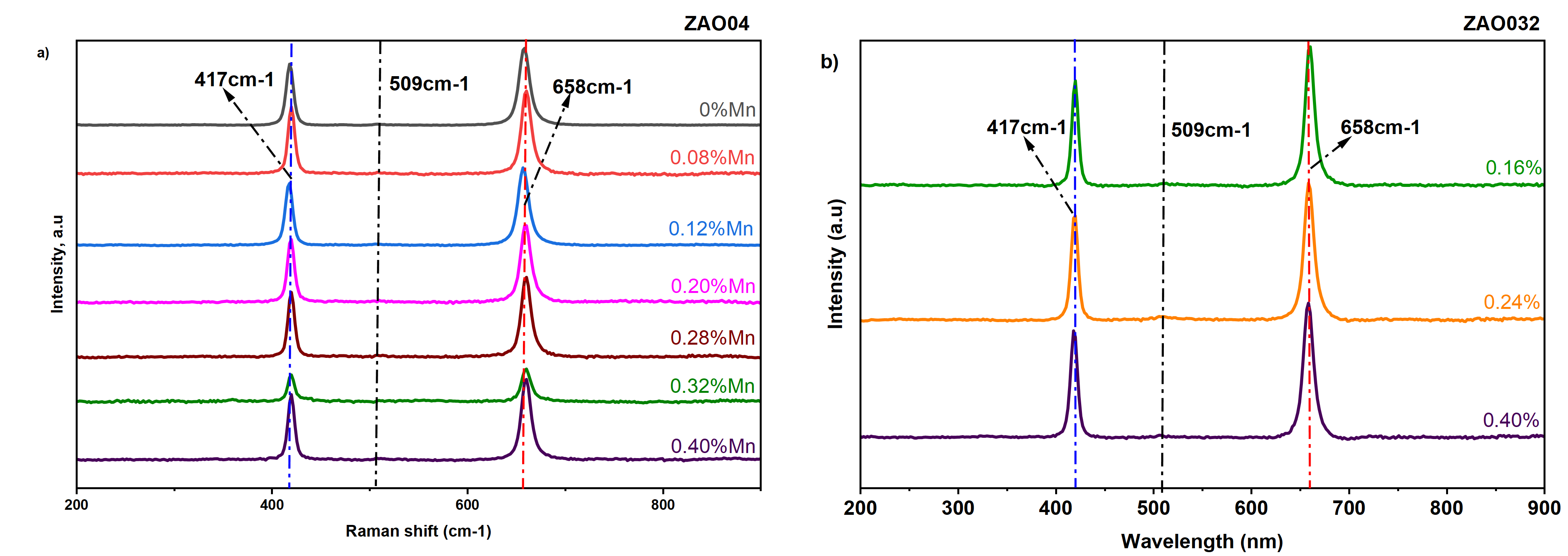}

\caption{Raman vibrational spectra of ZAO\_0.04\%Cr\_$y$\%Mn (a) and ZAO\_0.32\%Cr\_$y$\%Mn (b) sintered at 1100~$^{\circ}$C.}
\label{fig:S1}
\end{figure}

\begin{figure}[htbp]
\centering
\includegraphics[width=\columnwidth]{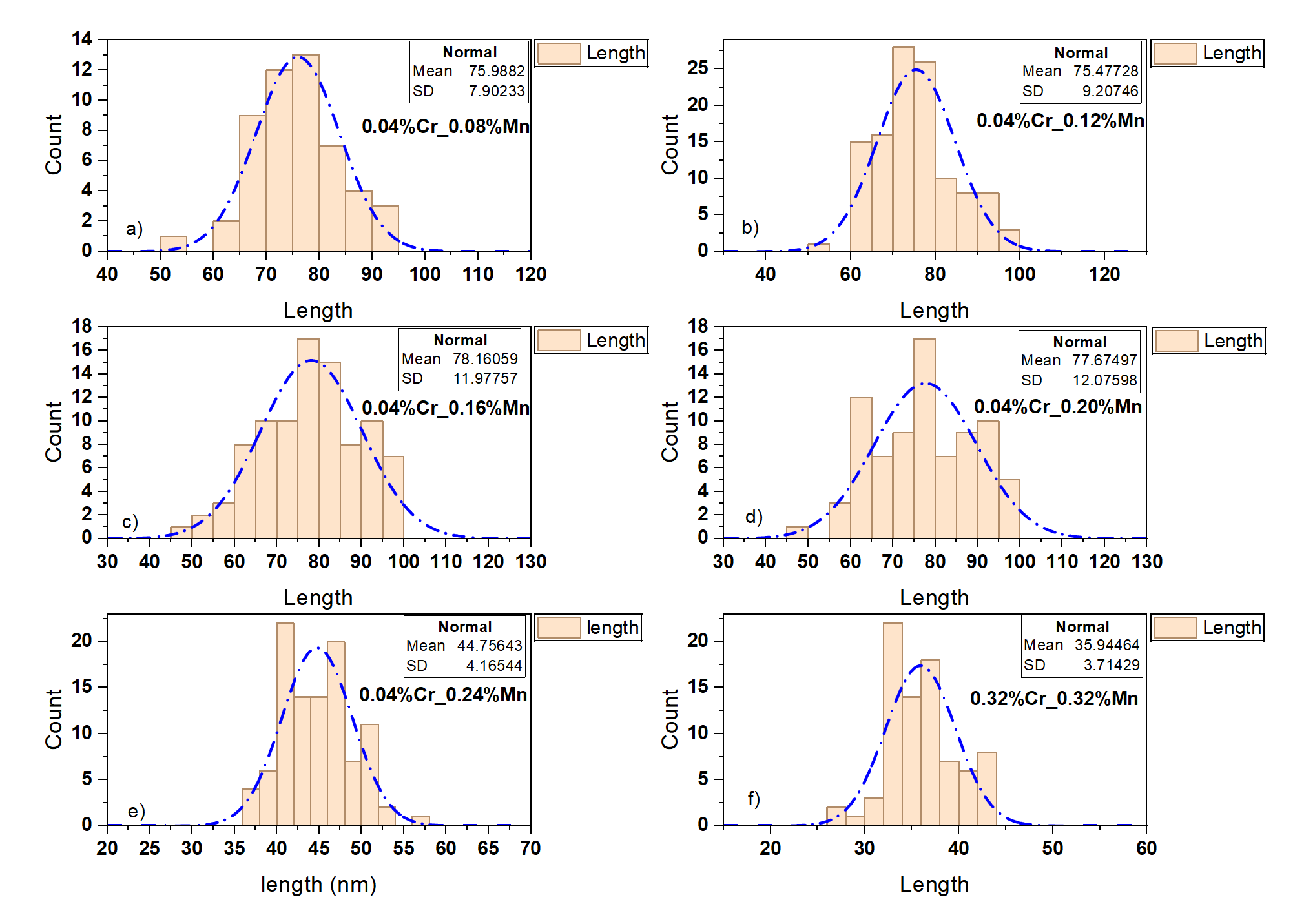}

\caption{Particle size distribution of ZAO4\_0.08, 0.12, 0.16, 0.20, and 0.24\%Mn (a--e) and ZAO32\_0.32\%Mn (f).}
\label{fig:S2}
\end{figure}

\begin{figure}[htbp]
\centering
\includegraphics[width=\columnwidth]{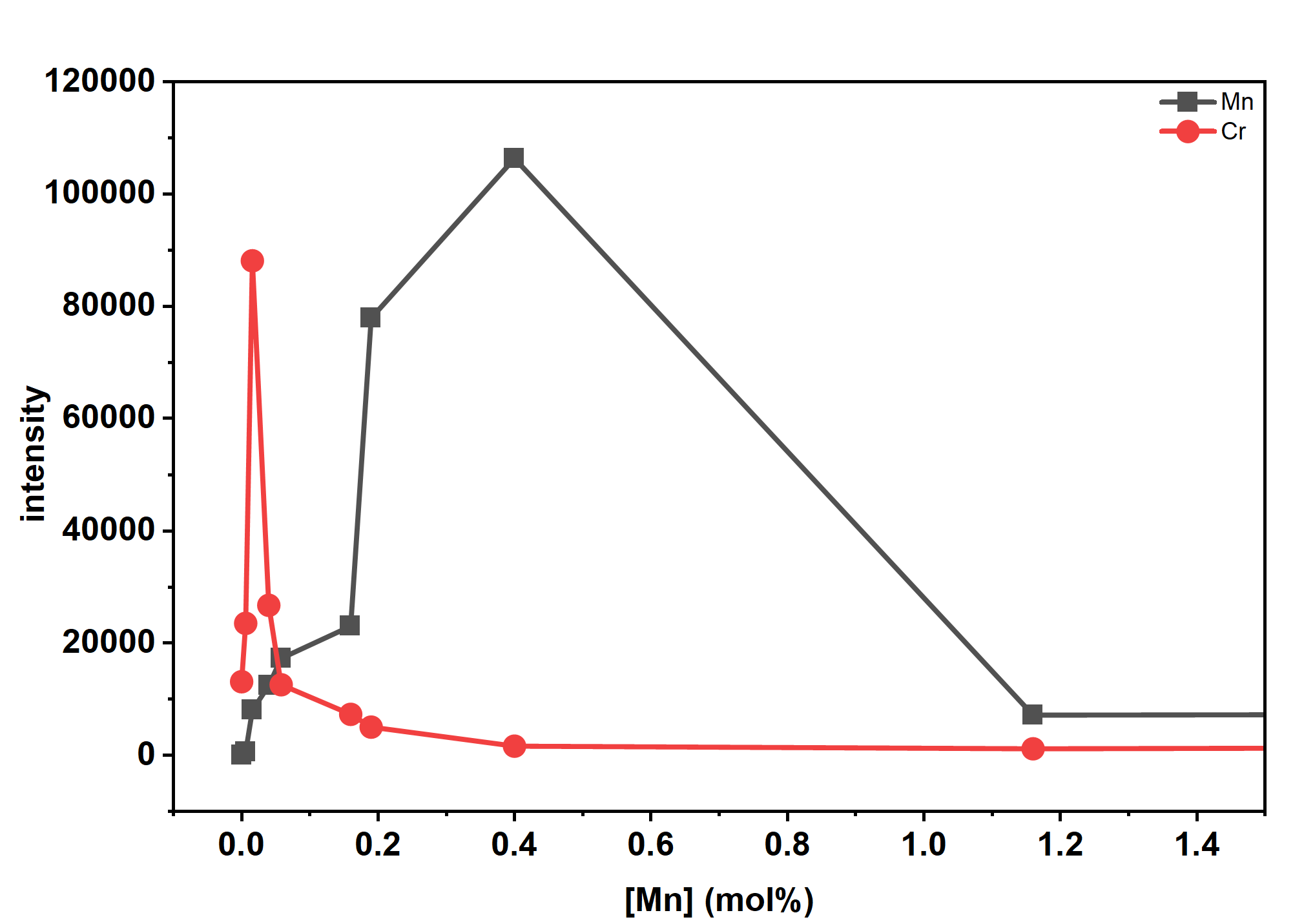}

\caption{Integrated photoluminescence intensities of the Mn-only reference series (no intentional Cr doping; 457~nm excitation): the \ch{Mn^2+} green band (470--620~nm, black squares) and the trace-\ch{Cr^3+} R/N-line region (650--740~nm, red circles) for the two measurement series (open symbols: 600 g/mm grating; filled symbols: 1800 g/mm grating), each scaled by the \ch{Cr^3+} intensity of its undoped member. The intensities are semi-quantitative: repeated measurements of one specimen differ by up to a factor of seven in absolute area, while their Cr/Mn ratio changes by only $\approx$20\%.}
\label{fig:S3}
\end{figure}

\begin{figure}[htbp]
\centering
\includegraphics[width=\columnwidth]{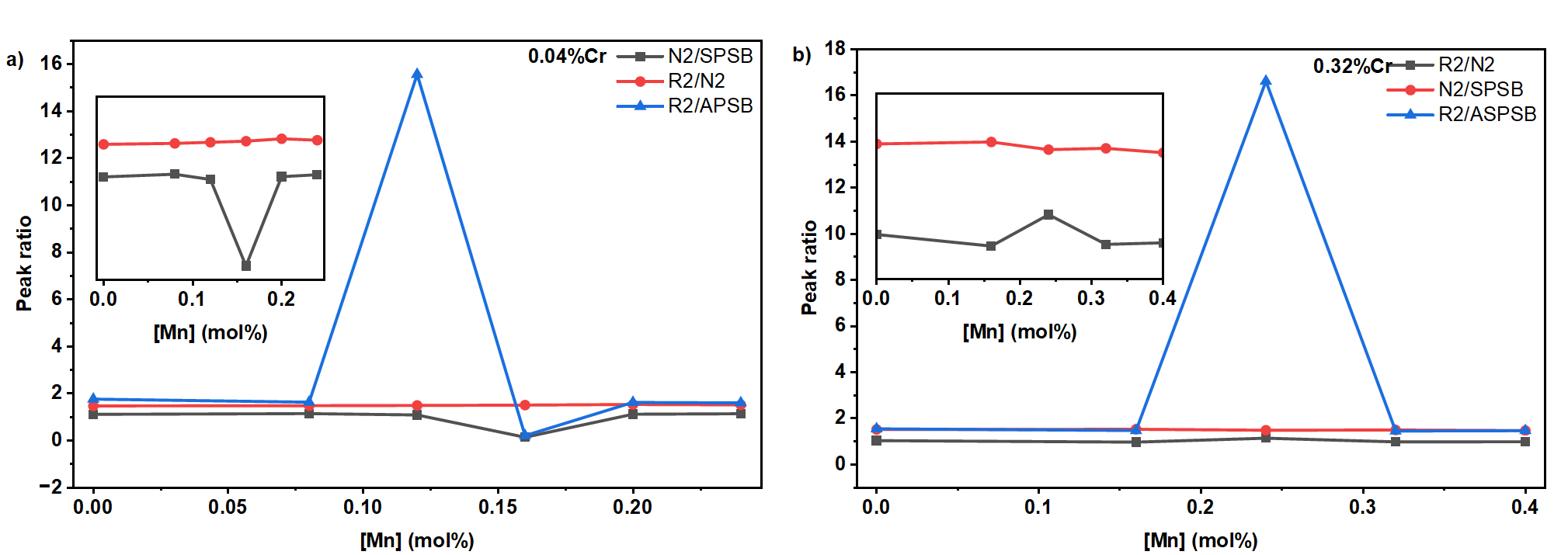}

\caption{Peak ratios of \ch{Cr^3+} emission as a function of \ch{Mn^2+} concentration: (a) ZAO4 and (b) ZAO32.}
\label{fig:S4}
\end{figure}

\begin{figure}[htbp]
\centering
\includegraphics[width=\columnwidth]{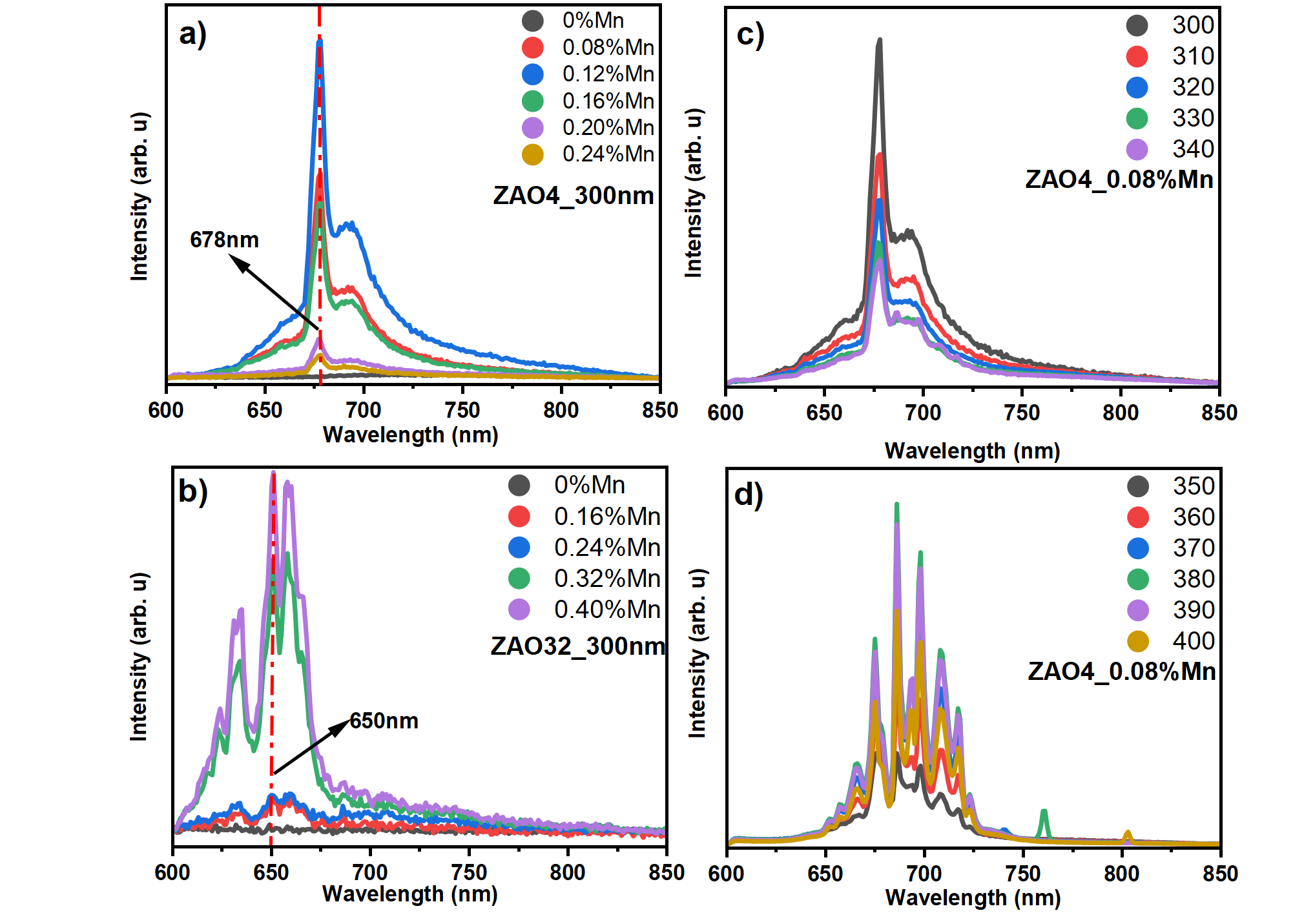}

\caption{PL emission of ZAO4 (a) and ZAO32 (b) under 300~nm excitation, and excitation-dependent emission of ZAO4\_0.08\%Mn at (c) 300--340~nm and (d) 350--400~nm.}
\label{fig:S5}
\end{figure}

\begin{figure}[htbp]
\centering
\includegraphics[width=\columnwidth]{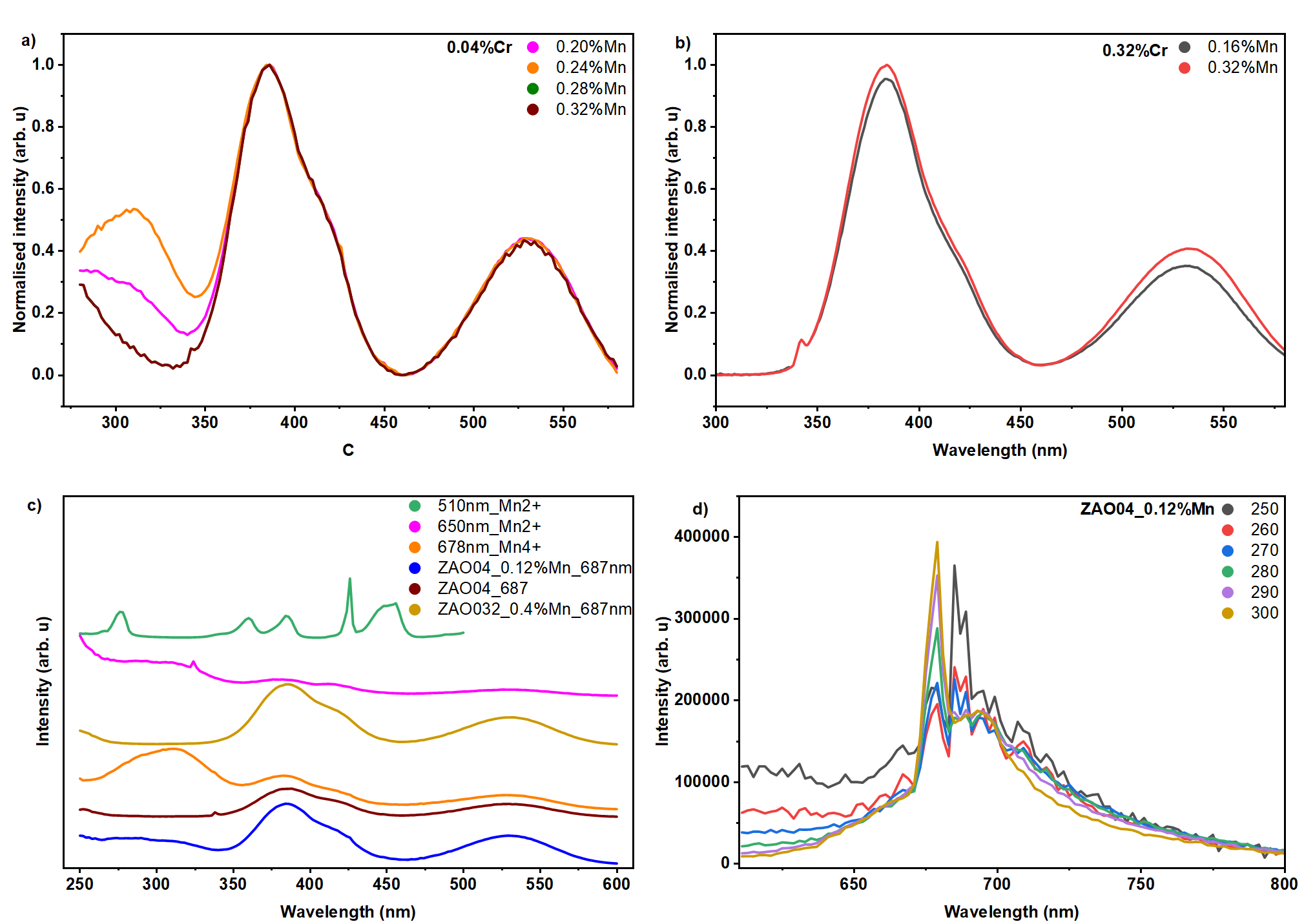}

\caption{(a) and (b) present the Cr PLE monitored at 687~nm for the ZAO4 and ZAO32 samples, respectively; (c) side-by-side plot of the Cr PLE and the Mn PLE monitored at 510~nm (\ch{Mn_{Tet}^{2+}}), 650~nm (\ch{Mn_{Oct}^{2+}}), and 678~nm (\ch{Mn_{Oct}^{4+}}), showing the spectral overlap between the Cr absorption bands and the Mn absorption bands monitored at 650 and 678~nm. Panel (d) shows the PL of the ZAO4\_0.12\%Mn sample in the region 250--3300~nm showing the gradual evolution of \ch{Mn^4+} emission at 678~nm.}
\label{fig:S6}
\end{figure}

The following equations were used to calculate the crystal field splitting parameters:
\begin{equation}
\frac{Dq}{B} = \frac{15(x - 8)}{x^{2} - 10x}
\label{eq:S1}
\end{equation}
\begin{equation}
\Delta E = \left({}^{4}A_{2} - {}^{4}T_{1}\right) - \left({}^{4}A_{2} - {}^{4}T_{2}\right)
\label{eq:S2}
\end{equation}
\begin{equation}
Dq = \left({}^{4}A_{2} - {}^{4}T_{2}\right)/10
\label{eq:S3}
\end{equation}
\begin{equation}
x = \frac{\Delta E}{Dq}
\label{eq:S4}
\end{equation}
\begin{equation}
B = \frac{2\nu_{1}^{2} + \nu_{2}^{2} - 3\nu_{1}\nu_{2}}{15\nu_{2} - 27\nu_{1}}
\label{eq:S5}
\end{equation}
where $B$ is the Racah parameter, and $\nu_{1}$ and $\nu_{2}$ are the energies of the $^{4}A_{2} \rightarrow {}^{4}T_{2}$ and $^{4}A_{2} \rightarrow {}^{4}T_{1}$ transitions, respectively.

\begin{table*}[htbp]
\centering
\caption{Crystal-field parameters of the Mn/Cr co-doped ZAO samples.}
\label{tab:S3}
\small
\begin{tabular}{@{}lcccccc@{}}
\toprule
Sample & $E_{1}$ ($^{4}A_{2}$--$^{4}T_{1}$) & $E_{2}$ ($^{4}A_{2}$--$^{4}T_{2}$) & $\Delta E$ & $B$ & $Dq$ & $Dq/B$ \\
 & (cm$^{-1}$) & (cm$^{-1}$) & (cm$^{-1}$) & (cm$^{-1}$) & (cm$^{-1}$) & \\
\midrule
\multicolumn{7}{l}{\textit{0.04\% Cr}} \\
0\%Mn & 26670.07 & 18589.73 & 8080.336 & 816.4961 & 1858.973 & 2.230076 \\
0.08\%Mn & 26112.99 & 18799.39 & 7313.6 & 724.7625 & 1879.939 & 2.593538 \\
0.12\%Mn & 26181.35 & 18799.39 & 7381.959 & 733.6745 & 1879.939 & 2.562034 \\
0.16\%Mn & 26112.99 & 18799.39 & 7313.6 & 724.7625 & 1879.939 & 2.593538 \\
0.20\%Mn & 25910.04 & 18941.81 & 6968.23 & 679.469 & 1894.181 & 2.787382 \\
0.24\%Mn & 25910.04 & 18870.33 & 7039.708 & 689.0746 & 1887.033 & 2.738154 \\
0.28\%Mn & 25910.04 & 18728.98 & 7181.059 & 679.469 & 1872.898 & 2.643175 \\
0.32\%Mn & 25910.04 & 18941.81 & 6968.23 & 679.469 & 1894.181 & 2.787382 \\
0.40\%Mn & 26044.99 & 18870.33 & 7174.656 & 706.1003 & 1887.033 & 2.672131 \\
\multicolumn{7}{l}{\textit{0.32\% Cr}} \\
0\%Mn & 26670.07 & 18589.73 & 8080.336 & 816.4961 & 1858.973 & 2.230076 \\
0.08\%Mn & 26388.59 & 18764.12 & 7624.47 & 766.4375 & 1876.412 & 2.447913 \\
0.16\%Mn & 26044.99 & 18834.79 & 7210.194 & 711.0271 & 1883.479 & 2.648618 \\
0.24\%Mn & 26112.99 & 18764.12 & 7348.871 & 729.7821 & 1876.412 & 2.570867 \\
0.32\%Mn & 26044.99 & 18799.39 & 7245.598 & 715.9728 & 1879.939 & 2.625378 \\
0.40\%Mn & 26112.99 & 18834.79 & 7278.196 & 719.7635 & 1883.479 & 2.616469 \\
\bottomrule
\end{tabular}
\end{table*}

\begin{table*}[htbp]
\centering
\caption{Integrated intensities, effective $g$ values, and peak-to-peak linewidths ($\Delta B_{\mathrm{pp}}$) of the low-field Cr signal of the ZAO4 and ZAO32 series.}
\label{tab:S4}
\small
\begin{tabular}{@{}lcccccc@{}}
\toprule
 & \multicolumn{2}{c}{Integrated intensity} & \multicolumn{2}{c}{$g_{\mathrm{eff}}$} & \multicolumn{2}{c}{$\Delta B_{\mathrm{pp}}$ (G)} \\
\cmidrule(lr){2-3}\cmidrule(lr){4-5}\cmidrule(lr){6-7}
{[}Mn{]} (mol\%) & ZAO4 & ZAO32 & ZAO4 & ZAO32 & ZAO4 & ZAO32 \\
\midrule
0 & 1.192 $\pm$ 0.038 & 1.7476 $\pm$ 0.010 & 3.8984 & 3.9159 & 30.49927 & 20.94553 \\
0.08 & 0.334 $\pm$ 0.035 & 0.627 $\pm$ 0.017 & 3.9065 & 3.9211 & 26.83936 & 20.33476 \\
0.12 & 0.257 $\pm$ 0.063 & -- & 3.8732 & -- & 26.83936 & -- \\
0.16 & 0.310 $\pm$ 0.052 & 0.582 $\pm$ 0.005 & 3.8695 & 3.9168 & 29.46927 & 20.66399 \\
0.20 & 0.231 $\pm$ 0.034 & -- & 3.8889 & -- & 25.61938 & -- \\
0.24 & 0.470 $\pm$ 0.307 & 0.734 $\pm$ 0.006 & 3.8811 & 3.9210 & 28.05933 & 20.94721 \\
0.28 & 0.429 $\pm$ 0.072 & -- & 3.8983 & -- & 25.61938 & -- \\
0.32 & 0.122 $\pm$ 0.062 & 0.639 $\pm$ 0.008 & 3.8090 & 3.9222 & 24.39941 & 20.13343 \\
0.40 & 0.255 $\pm$ 0.069 & 0.697 $\pm$ 0.009 & 3.8887 & 3.9234 & 26.83936 & 20.147833 \\
\bottomrule
\end{tabular}
\end{table*}

\begin{figure}[htbp]
\centering
\includegraphics[width=\columnwidth]{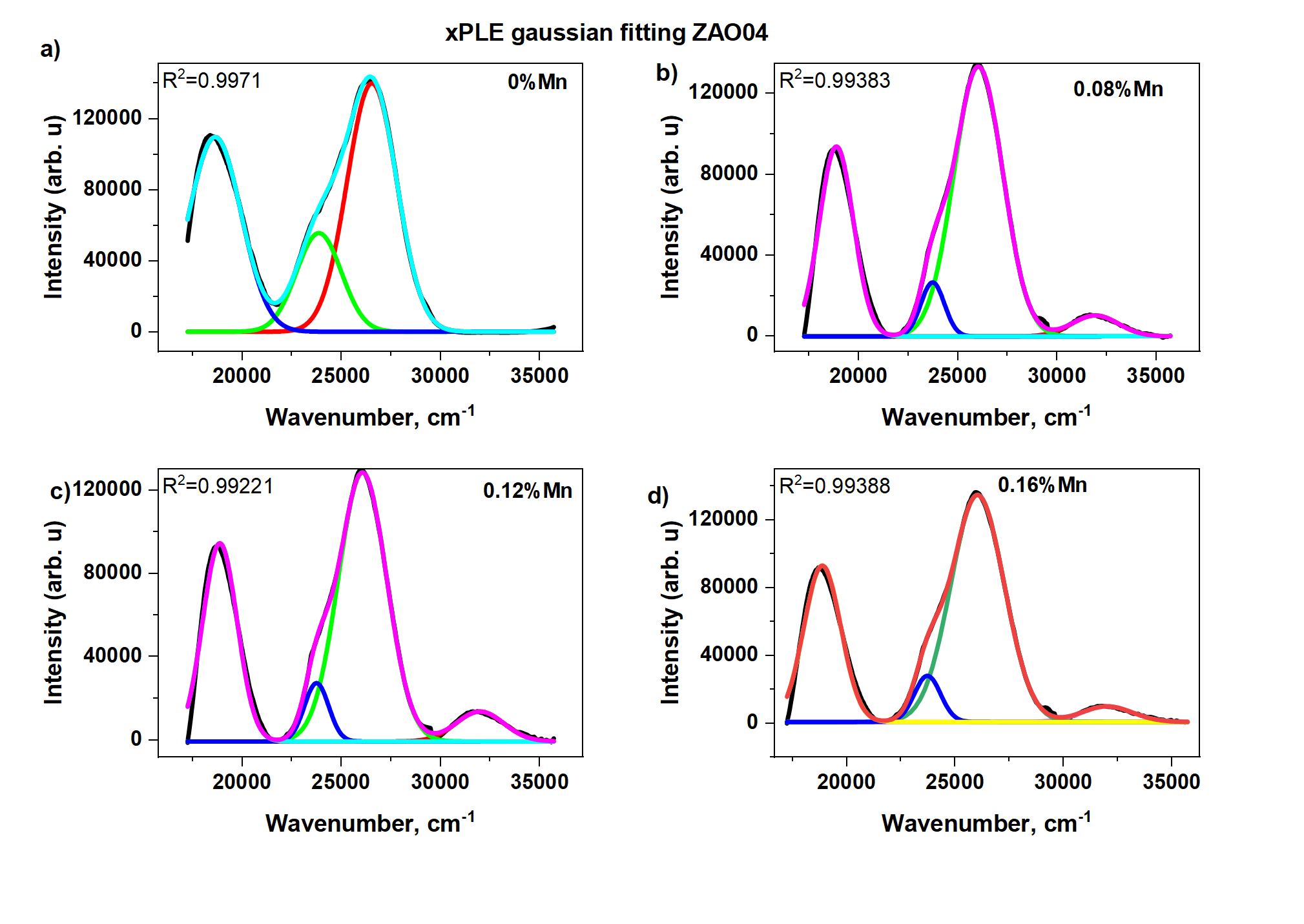}

\caption{(a--d) Gaussian decomposition of the PLE spectra of selected ZAO4 samples.}
\label{fig:S7}
\end{figure}

\begin{figure}[htbp]
\centering
\includegraphics[width=\columnwidth]{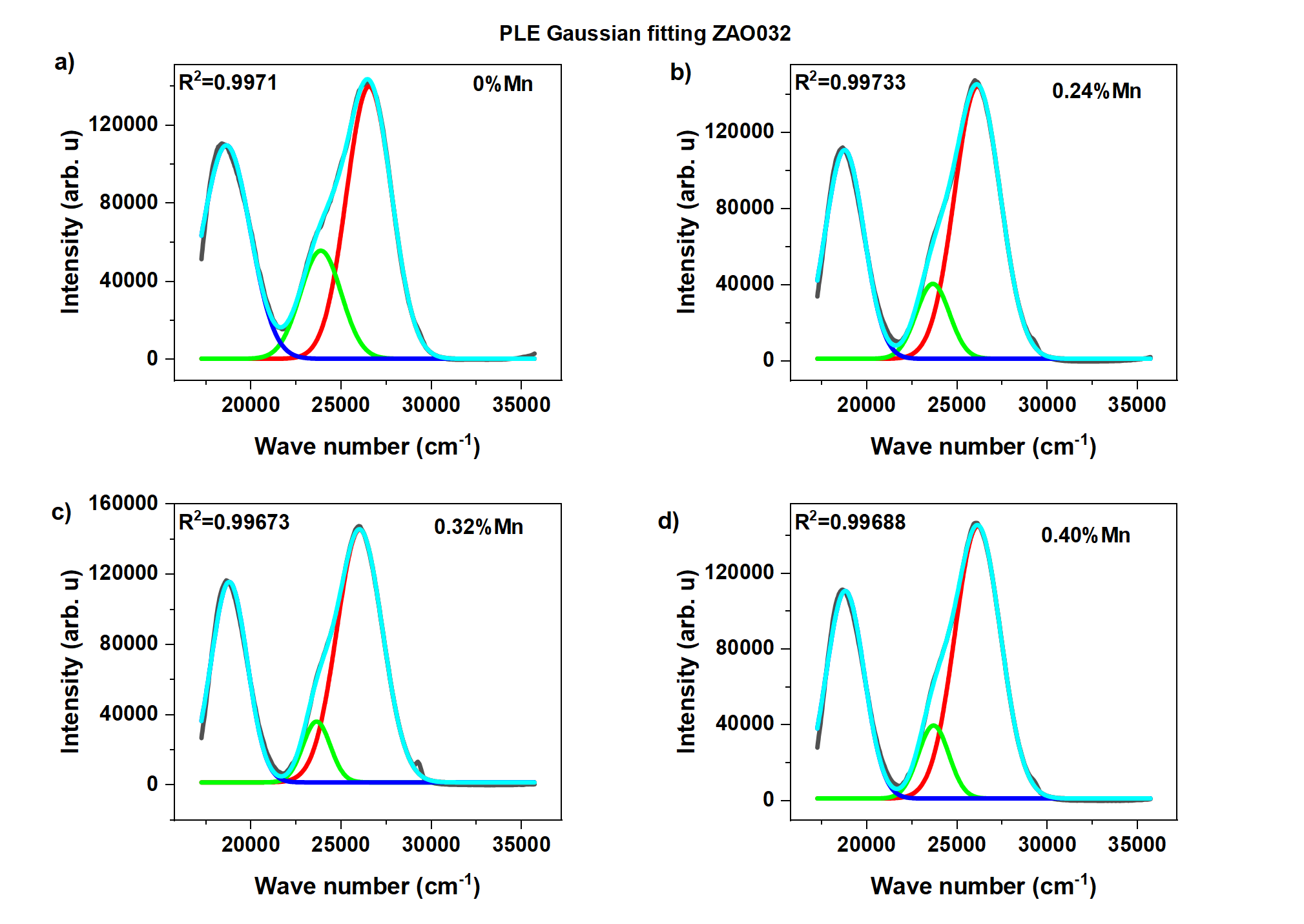}

\caption{(a--d) Gaussian decomposition of the PLE spectra of the ZAO32 samples.}
\label{fig:S8}
\end{figure}

\begin{table*}[htbp]
\centering
\caption{Spearman rank correlation coefficients $r_{\mathrm{s}}$ between the integrated spectroscopic observables of the ZAO4 series ($y$ = 0--0.40\% Mn). Each entry gives $r_{\mathrm{s}}$ with the two-sided permutation p-value ($2 \times 10^{4}$ permutations) and the number of common compositions $n$ in parentheses. The \ch{Mn^2+} EPR intensities are inverse-variance-weighted means over the three lowest microwave powers (0.10--1.0~mW), normalized per unit sample mass; the \ch{Cr^3+} EPR intensities are the double integrals of Table~\ref{tab:S4}. After Holm correction over the nineteen tests of Tables~\ref{tab:S5} and \ref{tab:S6}, only the \ch{Cr^3+} PL--XEOL pair remains individually significant (adjusted p $<$ 0.001; all other adjusted p $\geq$ 0.29, Benjamini--Hochberg q $\geq$ 0.15); the unadjusted values are therefore read as descriptive, and the model-level tests are collected in Table~\ref{tab:S8}. Restricted to the pre-collapse range ($y \leq 0.24\%$, n = 6): \ch{Cr^3+} PL--\ch{Mn^4+} $r_{\mathrm{s}}$ = +0.66; \ch{Cr^3+} PL--\ch{Mn^2+} EPR +0.77; \ch{Cr^3+} PL--510~nm \ch{Mn^2+} PL +0.20.}
\label{tab:S5}
\small
\begin{tabular}{@{}lccccc@{}}
\toprule
 & \textbf{\ch{Cr^3+} EPR} & \textbf{\ch{Cr^3+} PL} & \textbf{XEOL} & \textbf{\ch{Mn^2+} PL (510~nm)} & \textbf{\ch{Mn^4+} 678~nm} \\
\midrule
\textbf{\ch{Mn^2+} EPR} & $-$0.27 (0.50; 9) & +0.77 (0.023; 9) & +0.70 (0.043; 9) & +0.75 (0.025; 9) & +0.54 (0.30; 6) \\
\textbf{\ch{Cr^3+} EPR} & -- & $-$0.25 (0.52; 9) & $-$0.25 (0.52; 9) & 0.00 (1.00; 9) & $-$0.60 (0.24; 6) \\
\textbf{\ch{Cr^3+} PL} & -- & -- & +0.97 ($<$0.001; 9) & +0.68 (0.050; 9) & +0.66 (0.17; 6) \\
\textbf{XEOL} & -- & -- & -- & +0.65 (0.066; 9) & +0.66 (0.18; 6) \\
\textbf{\ch{Mn^2+} PL (510~nm)} & -- & -- & -- & -- & $-$0.14 (0.81; 6) \\
\bottomrule
\end{tabular}
\end{table*}

\begin{table}[htbp]
\centering
\caption{Spearman rank correlations for the ZAO32 series (same convention as Table~\ref{tab:S5}; \ch{Cr^3+} EPR intensities from Table~\ref{tab:S4}). The Holm and Benjamini--Hochberg corrections quoted in the caption of Table~\ref{tab:S5} cover these four tests as well.}
\label{tab:S6}
\small
\begin{tabular}{lccc}
\toprule
\textbf{Observable pair} & $r_{\mathrm{s}}$ & \textbf{p} & \textbf{n} \\
\midrule
\ch{Cr^3+} PL -- 650~nm octahedral \ch{Mn^2+} & +0.10 & 0.95 & 5 \\
\ch{Cr^3+} PL -- XEOL & +0.77 & 0.10 & 6 \\
Free-\ch{Cr^3+} EPR -- \ch{Cr^3+} PL & +0.14 & 0.81 & 6 \\
650~nm octahedral \ch{Mn^2+} -- [Mn] & +1.00 & 0.016 & 5 \\
\bottomrule
\end{tabular}
\end{table}

\begin{table}[htbp]
\centering
\caption{Principal-component analysis of the standardized ZAO4 observable matrix (six compositions, $y \leq 0.24\%$; six observables), reported as an exploratory summary: six observations cannot determine six-variable loadings precisely, and no significance is attached to them. PC1 and PC2 together capture 85\% of the variance: PC1 carries the associate channel (\ch{Cr^3+} PL, XEOL and \ch{Mn^4+} rising together with the free-\ch{Mn^2+} EPR signal, anticorrelated with the free-\ch{Cr^3+} EPR line) and PC2 the free-Mn channel (510~nm \ch{Mn^2+} emission). Under leave-one-out resampling the PC1 explained variance spans 56--67\% and the signs of its five non-negligible loadings are preserved in every subset, while the 510~nm \ch{Mn^2+} PL loading is indeterminate on PC1 and defines PC2 instead.}
\label{tab:S7}
\small
\begin{tabular}{lcc}
\toprule
 & \textbf{PC1} & \textbf{PC2} \\
\midrule
\textbf{Explained variance} & 60.8\% & 23.9\% \\
\ch{Mn^2+} EPR & +0.42 & $-$0.27 \\
\ch{Cr^3+} EPR (free) & $-$0.45 & +0.26 \\
\ch{Cr^3+} PL & +0.48 & +0.12 \\
XEOL & +0.49 & +0.11 \\
\ch{Mn^2+} PL (510~nm) & +0.09 & $-$0.77 \\
\ch{Mn^4+} 678~nm & +0.38 & +0.49 \\
\bottomrule
\end{tabular}
\end{table}

\begin{table}[htbp]
\centering
\caption{Spearman partial correlations controlling for the Mn content (ZAO4 series) and the global model-pattern tests. Partial correlations were computed on rank-transformed variables by regressing out the ranked Mn content and correlating the residuals, with two-sided p-values from $2 \times 10^{4}$ permutations of the residuals. The pattern statistic is the mean $r_{\mathrm{s}}$ over the pairs the association model predicts to co-vary positively --- a single test per series, so no multiplicity correction is involved --- with a one-sided p-value from $2 \times 10^{4}$ independent permutations of each observable.}
\label{tab:S8}
\small
\begin{tabular}{lccc}
\toprule
\textbf{Test} & $r_{\mathrm{s}}$ & \textbf{p} & \textbf{n} \\
\midrule
\ch{Cr^3+} PL -- XEOL & +0.97 & $< 10^{-4}$ & 9 \\
\ch{Mn^2+} EPR -- \ch{Cr^3+} PL & +0.77 & 0.015 & 9 \\
\ch{Mn^2+} EPR -- 510~nm \ch{Mn^2+} PL & +0.75 & 0.021 & 9 \\
\ch{Cr^3+} PL -- 510~nm \ch{Mn^2+} PL & +0.68 & 0.043 & 9 \\
\ch{Cr^3+} PL -- \ch{Mn^4+} 678~nm & +0.77 & 0.075 & 6 \\
Free-\ch{Cr^3+} EPR -- \ch{Cr^3+} PL & $-$0.32 & 0.39 & 9 \\
Pattern test, ZAO4 (mean $r_{\mathrm{s}}$, 4 pairs) & +0.76 & $< 10^{-4}$ & 9 \\
Pattern test, ZAO32 (mean $r_{\mathrm{s}}$, 2 pairs) & +0.89 & $1.1 \times 10^{-3}$ & 5--6 \\
\bottomrule
\end{tabular}
\end{table}

\end{document}